\documentclass[final,5p,times,twocolumn,authoryear]{elsarticle}

\usepackage{hyperref}
\hypersetup{colorlinks,linkcolor={blue},citecolor={blue},urlcolor={red}} 

\usepackage{color}

\newcommand{\vk}{\mathbf{k}}
\newcommand{\bs}{\mathbf{s}}
\newcommand{\bS}{\mathbf{S}}
\newcommand{\bd}{\mathbf{d}}
\usepackage{fontawesome}
\newcommand{\github}{\href{https://github.com/modichirag/flowpm}{\faGithub}}

\usepackage{aas_macros}

\usepackage{amssymb}

\usepackage{amsmath}
\usepackage{algorithm}
\usepackage{algpseudocode}
\algdef{SE}[VARIABLES]{Variables}{EndVariables}
   {\algorithmicvariables}
   {\algorithmicend\ \algorithmicvariables}
\algnewcommand{\algorithmicvariables}{\textbf{global variables}}

\usepackage{listings}

\usepackage[frozencache]{minted}
\usepackage{color}

\definecolor{mygreen}{rgb}{0,0.6,0}
\definecolor{mygray}{rgb}{0.5,0.5,0.5}
\definecolor{mymauve}{rgb}{0.58,0,0.82}

\lstnewenvironment{code}[1][]%
  {\noindent\minipage{\linewidth}\medskip 
   \lstset{basicstyle=\ttfamily\footnotesize,frame=single,#1}}
  {\endminipage}

\journal{Astronomy and Computing}

\begin{document}

\begin{frontmatter}




\title{\texttt{FlowPM}: Distributed TensorFlow Implementation of the \texttt{FastPM} Cosmological N-body Solver}


\author[bccp]{Chirag Modi}
\author[cea]{Fran\c{c}ois Lanusse}
\author[bccp,lbl]{Uro\v{s} Seljak}
\address[bccp]{Berkeley Center for Cosmological Physics, Department of Physics, University of California, Berkeley, CA 94720, USA}
\address[cea]{AIM, CEA, CNRS, Universit\'e Paris-Saclay, Universit\'e Paris Diderot, Sorbonne Paris Cit\'e, F-91191 Gif-sur-Yvette, France}
\address[lbl]{Lawrence Berkeley National Laboratory, One Cyclotron Road, Berkeley, CA 94720, USA}

\begin{abstract}

We present \texttt{FlowPM}, a Particle-Mesh (PM) cosmological N-body code implemented in Mesh-TensorFlow for GPU-accelerated, distributed, and differentiable simulations.
We implement and validate the accuracy of a novel multi-grid scheme based on multiresolution pyramids to compute large scale forces efficiently on distributed platforms.
We explore the scaling of the simulation on large-scale supercomputers and compare it with corresponding python based PM code, finding on an average 10x speed-up in terms of wallclock time.
We also demonstrate how this novel tool can be used for efficiently solving large scale cosmological inference problems, in particular reconstruction of cosmological fields in a forward model Bayesian framework with hybrid PM and neural network forward model.
We provide skeleton code for these examples and the entire code is publicly available at \href{ }{https://github.com/modichirag/flowpm}. \github
\end{abstract}




\end{frontmatter}


\section{Introduction}
\label{sec:intro}

N-Body simulations evolving billions of dark matter particles under gravitational forces have become essential tool for modern day cosmology data analysis and 
interpretation.
They are required for accurate predictions of the large scale structure (LSS) of the Universe, probed by tracers such as galaxies, quasars, gas, weak gravitational lensing etc.
While full N-body simulations can provide very precise predictions, they are extremely slow and computationally expensive.
As a result they are usually complemented by fast approximated numerical schemes such as Particle-Mesh (PM) solvers \citep{Merz05, Tassev13, White14, Feng2016, Izard16}.
In these approaches, the gravitational forces experienced by dark matter particles in the simulation volume are computed by Fast Fourier Transforms on a 3D grid.
This makes PM simulations very computationally efficient and scalable, at the cost of approximating the interactions on small scales close to the resolution of the grid and ignoring the force calculation on sub-resolution scales.

The next generation of cosmological surveys such as Dark Energy Spectroscopic Instrument (DESI) \citep{DESI}, the Rubin Observatory Legacy Survey of Space and Time (LSST) \citep{LSST}, and others will probe LSS with multiple tracers over unprecedented scales and dynamic range.
This will make both the modeling and the analysis challenging even with current PM simulations.
To improve the modeling in terms of dynamic range and speed, many recent works have explored applications of machine learning, with techniques based on deep convolutional networks and generative models \citep{He19, Ramanah20}.
However these approaches often only learn the mapping between input features (generally initial density field) and final observables and completely replace the correct underlying physical dynamics of the evolution with end-to-end modeling.
As a result, the generalization of these models across redshifts (time scales), length scales (cosmological volume, resolution etc.), cosmologies and observables of interest is limited by the training dataset and needs to be explored exhaustively to account for all failure modes. 

At the same time, there has been a considerable interest in exploring novel techniques to push simulations to increasingly smaller and non-Gaussian regimes for the purpose of cosmological analyses.
Recent work has demonstrated that one of the most promising approaches to do this is with forward modeling frameworks and using techniques like simulations based inference or reconstruction of cosmological fields for analysis \citep{Seljak17, Alsing18, Cranmer19}.
However given the highly non-linear forward dynamics and multi-million dimensionality of the cosmological simulations, such approaches are only tractable if one has access to differentiable forward simulations \citep{Jasche13, Wang14, Seljak17, Modi18}. 

To address both of these challenges, in this work we present \texttt{FlowPM}, a TensorFlow implementation of an N-Body PM solver.
Written entirely in TensorFlow, these simulations are GPU-accelerated and hence faster than CPU-based PM simulations, while also being entirely differentiable end-to-end.
At the same time, they encode the exact underlying dynamics for gravitational evolution (within a PM scheme), while still leaving room for natural synthesis with machine learning components to develop hybrid simulations. These hybrids could use machine learning to model sub-grid and non-gravitational dynamics otherwise not included in a PM gravity solver. 

As they evolve billions of particles in order to simulate today's cosmological surveys, N-Body simulations are quite memory intensive and necessarily require distributed computing.
In \texttt{FlowPM}, we develop such distributed computation with a novel model parallelism framework - Mesh-TensorFlow \citep{Shazeer2018}.
This allows us to control distribution strategies and split tensors and computational graphs over multiple processors.

The primary bottleneck for large scale distributed PM simulations are distributed 3D-Fast Fourier Transforms (FFT), which incur large amounts of communications and  make up for more than half the wall-clock time \citep{HiddenValley19}.
One way to overcome these issues is with by using a multi-grid scheme for computing gravitational forces \citep{Suisalu95, Merz05, Harnois13}.
In these schemes, large scale forces are estimated on a coarse distributed grid and then stitched together with small scale force estimated locally.
Following the same idea, we also propose a novel multi-grid scheme based on Multiresolution Pyramids \citep{Burt83, Anderson84}, that does not require any fine-tuned stitching of scales, and benefits from highly optimized 3-D convolutions of TensorFlow.

In this first work, we have implemented the \texttt{FastPM} scheme of \cite{Feng2016} for PM evolution with force resolution $B=1$.
Different PM schemes, as well as other approximate simulations like Lagrangian perturbation theory fields \citep{Tassev13, Kitaura14, Stein19}, are built upon the same underlying components, with interpolations to-and-from PM grid and efficient FFTs to estimate gravitational forces being the key 
elements.
Thus modifying \texttt{FlowPM} to include other PM schemes is a matter of implementation rather than methodology development. 
Therefore in this work, we will focus in detail on these underlying components while only giving the outline of the full algorithm. 
For further technical details on the choices and accuracy of the implemented \texttt{FastPM} scheme, we refer the reader to the original paper of \cite{Feng2016}.

The structure of this paper is as follows - 
in Section \ref{sec:fastpm} we outline the basic \texttt{FastPM} algorithm and discuss our multi-grid scheme for force estimation in Section \ref{sec:multigrid}.
Then in Section \ref{sec:mesh}, we introduce Mesh-TensorFlow.
In Section \ref{sec:flowpm}, we build upon these and introduce \texttt{FlowPM}.
We compare the scaling and accuracy of \texttt{FlowPM} with the python \texttt{FastPM} code in Section \ref{sec:scaling}.
At last, in Section \ref{sec:reconstruction}, we demonstrate the efficacy of \texttt{FlowPM} with a toy example by combining it with neural network forward models and reconstructing the initial conditions of the Universe.
We end with discussion in Section \ref{sec:discussion}.

Throughout the paper, we use ``grid" to refer to the particle-mesh grid and ``mesh" to refer to the computational mesh of Mesh-TF.
Unless explicitly specified, every FFT referred throughout the paper is a 3D FFT.

\section{The \texttt{FastPM} particle mesh N-body solver}
\label{sec:fastpm}

Schematically, \texttt{FastPM} implements a symplectic series of Kick-Drift-Kick operations \citep{Quinn97} to do the time integration for gravitational evolution following leapfrog integration,
starting from the Gaussian initial conditions of the Universe to the observed large scale structures at the final time-step.
In the \textit{Kick} stage, we estimate the gravitational force on every particle and update their momentum $p$.
Next, in the staggered \textit{Drift} stage, we displace the particle to update their position ($x$) with the current velocity estimate.
The drift and kick operators can thus be defined as \citep{Quinn97} - 
\begin{align}
x(t_1) &= x(t_0) + p(t_0) \mathcal{D}_\mathrm{PM} \\
p(t_1) &= p(t_0) + f(t_0) \mathcal{K}_\mathrm{PM},
\end{align}
where $\mathcal{D}_\mathrm{PM}$ and $\mathcal{K}_\mathrm{PM}$ are scalar drift and kick factors and $f(t_r)$ is the gravitational force.
For the leapfrog integration implemented in \texttt{FlowPM} (and \texttt{FastPM}), these operators are modified to include the position and velocity at staggered half time-steps instead of the initial time ($t_0$) in kick and drift steps respectively.

Estimating the gravitational force is the most computationally intensive part of the simulation. 
In a PM solver, the gravitational force is calculated via 3D Fourier transforms. 
First, the particles are interpolated to a spatially uniform grid with a kernel $W(r)$ to estimate the mass overdensity field at every point in space. 
The most common choice for the kernel, and the one we implement, is a linear window of unite size corresponding to the cloud-in-cell (CIC) interpolation scheme \citep{Hockney88}.
We then apply a Fourier transform to obtain the over-density field $\delta_k$ in Fourier space. 
This field is related to the force field via a transfer function ($\nabla \nabla^{-2}$).
\begin{equation}
f(\vk) = \nabla\nabla^{-2} \delta(\vk)
\label{eq:forcek}
\end{equation}
Once the force field is estimated on the spatial grid, it is interpolated back to the position of every particle with the same kernel as was used to generate the density field in the first place. 

There are various ways to write down the transfer function in a discrete Fourier space, as explored in detail in \cite{Hockney88}.
The simplest way of doing this is with a naive Green's function kernels ($\nabla^{-2} =k^{-2}$) and differentiation kernels ($\nabla = i \bf{k}$). In \texttt{FlowPM} however, we implement a finite differentiation kernel as used in \texttt{FastPM} with
\begin{equation}
\nabla^{-2} = \Bigg( \sum_{d=x,y,z} \Big(x_0\omega_0 {\rm sinc} \frac{\omega_d}{2} \Big)^2 \Bigg)^{-1}
\end{equation}{}
\begin{equation}
\nabla = D_1(\omega) = \frac{1}{6} \Big( 8{\rm sin}\omega - {\rm sin}2\omega )
\end{equation}{}
where $x_0$ is the grid size and $\omega = k x_0$ is the circular frequency that goes from $(-\pi, \pi]$.

\section{Multi-grid PM Scheme}
\label{sec:multigrid}

Every step in PM evolution requires one forward 3D FFT to estimate the overdensity field in Fourier space,
and three inverse FFTs to estimate the force component in each direction.
The simplest way to implement 3D FFTs is as 3 individual 1D Fourier transforms in each direction \citep{pfft}. 
However in a distributed implementation, where the tensor is distributed on different processes, this involves transpose operations and expensive all-to-all communications in order to get the dimension being transformed on the same process.
This makes these FFTs the most time-intensive step in the simulation \citep{HiddenValley19}.
There are several ways to make this force estimation more efficient, such as fast multipole methods \citep{Greengard87} and multi-grid schemes \citep{Merz05, Harnois13}.
In FlowPM, we implement a novel version of the latter. 

The basic idea of a multi-grid scheme is to use grids at different resolutions, with each of them estimating force on different scales which are then stitched together.
Here we discuss this approach for a 2 level multi-grid scheme, since that is implemented in \texttt{FlowPM} and the extension to more levels is similar. 
The long range (large-scale) forces are computed on a global coarse grid that is distributed across processes.
On the other hand, the small scale forces are computed on a fine, higher resolution grid that is local, i.e. it estimates short range forces on smaller independent sections that are hosted locally by individual processes.
Thus, the distributed 3D FFTs are computed only on the coarse mesh while the higher resolution mesh computes highly efficient local FFTs.

This force splitting massively reduces the communication data-volume but at the cost of increasing the total number of operations performed.
Thus while it scales better for large grids and highly distributed meshes, it can be excessive for small simulations.
Therefore in \texttt{FlowPM} we implement both, a multi-grid scheme for large simulations, and the usual single-grid force scheme for small grid sizes, with the user having the freedom to choose between the two.

\subsection{Multiresolution Pyramids}
\label{sec:pyramids}
In this section we briefly discuss multiresolution pyramids that will later form the basis our multi-grid implementation for force estimation.
As used in image processing and signal processing communities, image pyramids refer to multi-scale representations of signals and image built through recursive reduction operations on the original image \citep{Burt83, Anderson84}. 
The reduction operation is a combination of - i) smoothing operation, often implemented by convolving the original image with a low-pass filter ($\mathcal{G}$) to avoid aliasing effects by reducing the Nyquist frequency of the signal, and ii) subsampling or downsampling operation ($\mathcal{S}$), usually by a factor of 2 along each dimensions, to compress the data. 
This operation can be repeatedly applied at every `level' to generate a `pyramid' structure of the low-pass filtered copies of the original image.  
Given a level $g_{l}$ of original image $g_0$, the pyramid next level is generated by 
\begin{equation}
    g_{l+1}={\rm REDUCE}(g_{l}) = \mathcal{S}[\mathcal{G} \star g_l] 
\end{equation}
where $\star$ represents a convolution. 
If the filter $\mathcal{G}$ used is a Gaussian filter, the pyramid is called a Gaussian pyramid.

\begin{figure}[h]
\centering
\includegraphics[width=.5\textwidth]{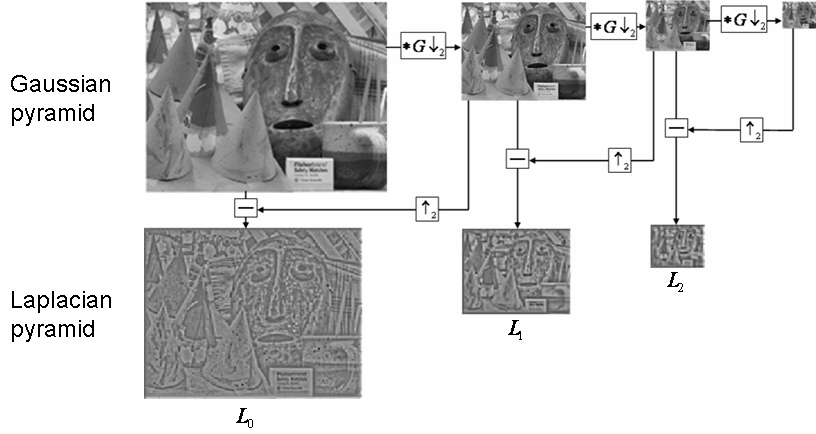}
\caption{An example of Gaussian image pyramid and the corresponding Laplacian pyramid. The arrows and small boxes mark the operations involved\protect \footnotemark. }
\label{fig:pyramid}
\end{figure}

However, in our multi-grid force estimation, we are interested in decoupling the large and small scale forces.
This decoupling can be achieved by building upon the multi-scale representation of Gaussian pyramids to construct independent band-pass filters with a `Laplacian' pyramid\footnote{Since we will not use the Gaussian kernel for smoothing, our pyramid structure is not a Laplacian pyramid in the strictest sense, but the underlying idea is the same.}.In this case, one reverse the reduce operation by - i) Upsampling ($\mathcal{U}$) the image at level $g_{l+1}$, generally by inserting zeros between pixels and then ii) interpolating the missing values by convolving it with the same filter $\mathcal{G}$ to create the expanded low-pass image $g'_l$ 
\begin{equation}
    g'_{l}={\rm EXPAND}(g_{l+1}) = \mathcal{G}\star \mathcal{U}[g_{l+1}] 
\end{equation}
A Laplacian representation at any level can then be constructed by taking the difference of the original and expanded low-pass filtered image at that level
\begin{equation}
    L_{l}= g_l - g'_l
\end{equation}
In Figure \ref{fig:pyramid}, we show the Gaussian and Laplacian pyramid constructed for an example image, along with the flow-chart for these operations.

\footnotetext{Image taken from \href{ }{http://graphics.cs.cmu.edu/courses/15-463/2012\_fall/hw/proj2g-eulerian/}}

\begin{figure}[h]
\centering
\includegraphics[width=.5\textwidth]{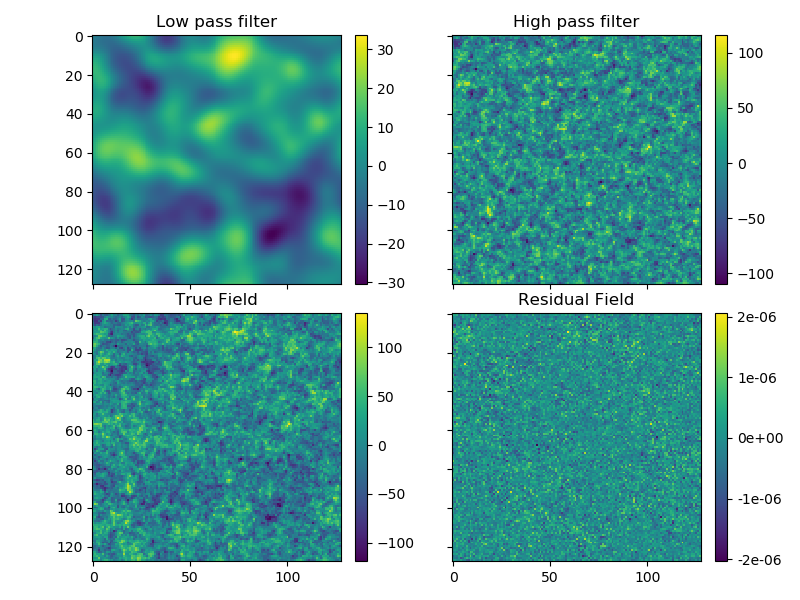}
\caption{Laplacian pyramid for matter density field - the top row represents the upsampled low-pass filtered field and the high-pass filter field generated by taking the difference with the true field (bottom left). Residuals (bottom right) demonstrate that the original field is reconstructed perfectly with low and high pass filter fields.}
\label{fig:levels}
\end{figure}
\footnotetext{Image taken from \href{ }{http://graphics.cs.cmu.edu/courses/15-463/2012\_fall/hw/proj2g-eulerian/}}

\begin{figure}[h]
\centering
\includegraphics[width=.5\textwidth]{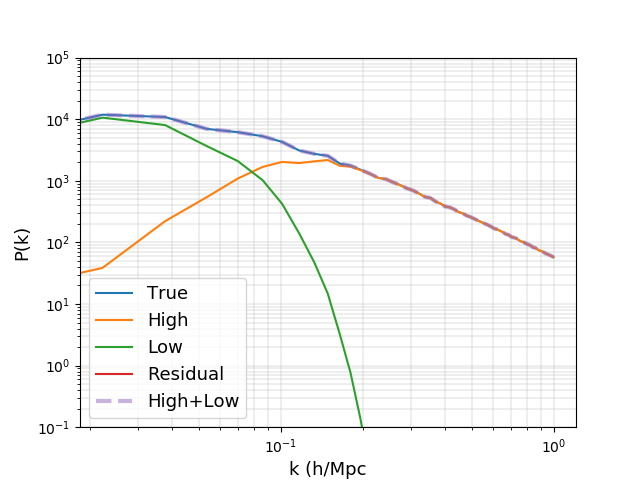}
\caption{The power spectrum corresponding to the density fields shown in Figure \ref{fig:levels}.}
\label{fig:plevels}
\end{figure}
\footnotetext{Image taken from \href{ }{http://graphics.cs.cmu.edu/courses/15-463/2012\_fall/hw/proj2g-eulerian/}}

Lastly, given the image at the highest level of the pyramid ($g_N$) and the Laplacian representation at every level ($L_0...L_{N-1}$), we can exactly recover the original image by recursively doing the series of following inverse transform- 
\begin{equation}
    g_{l}= L_l + {\rm EXPAND}(g_{l+1}) \quad \forall l \in [N-1...0]
\end{equation}

In Figure \ref{fig:levels}, we show the resulting fields when these operations are applied on the matter density field. We show the result of upsampling the low-pass filtered image and the high pass filtered image with the low-pass filtered component removed as is done in Laplacian pyramid. The negligible residuals show that the reconstruction of the fields is exact. In Figure \ref{fig:plevels}, we show the power spectra of the corresponding fields to demonstrate which scales are contributed at every level. We will apply these operations in Section \ref{sec:forcepyramids} to estimate forces on a two-level grid. 

\section{Mesh-TensorFlow}
\label{sec:mesh}

The last ingredient we need to build \texttt{FlowPM} is a model parallelism framework that is capable of distributing the N-body simulation over multiple processes.
To this end, we adopt Mesh-TensorFlow \citep{Shazeer2018} as the framework to implement our PM solver.
Mesh-TensorFlow is a framework for distributed deep learning capable of specifying a general class of distributed tensor computations.
In that spirit, it goes beyond the simple batch splitting of data parallelism and generalizes to be able to split any tensor and associated computations across any dimension on a mesh of processors. 

We only summarize the key component elements of Mesh-TensorFlow here and refer the reader to \cite{Shazeer2018} \footnote{\href{https://github.com/tensorflow/mesh}{https://github.com/tensorflow/mesh}} for more details.
These are - 
\begin{itemize}
    \item a \textit{mesh}, which is a $n$-dimensional array of processors with \textit{named} dimensions.
    \item the dimensions of every tensor are also \textit{named}. Depending on their name, these dimensions are either distributed/split across or replicated on all processors in a mesh. Hence every processor holds a \textit{slice} of every tensor.
    The dimensions can thus be seen as `logical' dimensions which will be split in the same manner for all tensors and operations. 
    \item a user specified \textit{computational layout} to map which tensor dimension is split along which mesh dimension.
    The unspecified tensor dimensions are replicated on all processes. 
    While the layout does not affect the results, it can affect the performance of the code.
\end{itemize}

The user builds a Mesh-TensorFlow graph of computations in Python and this graph is then \textit{lowered} to generate the TensorFlow graphs.
It is at this stage that the defined tensor dimensions and operations are split amongst different processes on the mesh based on the computational layout.
Multi-CPU/GPU meshes are implemented with \textit{PlacementMeshImpl}, i.e. a separate TensorFlow operations is placed on the different devices, all in one big TensorFlow graph.
In this case, the TensorFlow graph will grow with the number of processes.
On the other hand, TPU meshes are implemented in with \textit{SimdMeshImpl} wherein TensorFlow operations (and communication collectives) are emitted from the perspective of one core, and this same program runs on every core, relying on the fact that each core actually performs the same operations.
Due to Placement Mesh Implementation, time for lowering also increases as the number of processes are increased.
Thus currently, it is not infeasible to extend Mesh-TensorFlow code to a very large number of GPUs for very large computations.
We expect such bottlenecks to be overcome in the near future.
Returning to Mesh-TF computations, lastly, the tensors to be evaluated are \textit{exported} to TF tensors in which case they are gathered on the single process. 

\begin{figure}[h]
\begin{minted}[frame=single,
               fontsize=\footnotesize]{python}
import mesh_tensorflow as mtf
import tensorflow.compat.v1 as tf

# Setup graph and associated mesh
graph = mtf.Graph()
mesh = mtf.Mesh(graph, "my_mesh")

# Define named dimensions for defining a 3D volume
x_dim, y_dim, z_dim = (mtf.Dimension("nx", 128), 
                       mtf.Dimension("ny", 128), 
                       mtf.Dimension("nz", 128))
batch_dim = mtf.Dimension("batch", 1)

# Sample a batched random normal 3D volume
field = mtf.random_normal(mesh, 
        [batch_dim, x_dim, y_dim, z_dim])

# [...] Other mtf operations can be added here

# Define a concrete implementation as a 2D mesh on 8 GPUs,
# splitting `nx` and `ny` dimensions along rows and cols
mesh_impl = mtf.placement_mesh_impl.PlacementMeshImpl(
        mesh_shape=[("row", 4), ("col", 2)], 
        layout_rules=[("nx", "row"), ("ny", "col")], 
        devices=["device:GPU:%d"%i for i in range(8)])

#Lower the graph onto computational mesh
lowering = mtf.Lowering(graph, {mesh:mesh_impl})
# Retrieve mtf tensor as TensorFlow tensor
tf_field = lowering.export_to_tf_tensor(field)

# Evaluate graph
with tf.Session as sess: 
    sess.run(tf_field)
\end{minted}
\caption{Sample code to generate random normal field of grid size 128 on a 2D computational mesh of 8 GPUs with x and y directions split across 4 and 2 processors respectively}
\label{lst:meshexample}
\end{figure}

Figure \ref{lst:meshexample} shows an example code to illustrate these operations which form the starting elements of any Mesh-TensorFlow.
The code also sets up context for FlowPM, naming the three directions of the PM simulation grid and explicitly specifying which direction is split along which mesh-dimension.
Note that for a batch of simulations, computationally the most efficient splitting will almost always maximize the splits along the batch dimension.
While \texttt{FlowPM} supports that splitting, we will henceforth fix the batch size to 1 since our focus is on model parallelism of large scale simulations.

\section{\texttt{FlowPM}: \texttt{FastPM} implementation in Mesh-Tensorflow}
\label{sec:flowpm}

Finally, in this section, we introduce \texttt{FlowPM} and discuss technical details about its implementation in Mesh-TensorFlow.
We will not discuss the underlying algorithm of a PM simulation since it is similar in spirit to any PM code, and is exactly the same as \texttt{FastPM} \citep{Feng2016}.
Rather we will focus on domain decomposition and the multi-grid scheme for force estimation which are novel to FlowPM.

\subsection{Domain decomposition and Halo Exchange}
\label{sec:haloex}
In Mesh-Tensorflow, every process has a slice of every tensor.
Thus for our PM grid, where we split the underlying grid spatially along different dimensions, every process will hold only the physical region and the particles corresponding to that physical region.
However at every PM step, there is a possibility of particles moving in-and-out of any given slice of the physical region.
At the same time, convolution operations on any slice need access to the field outside the slice when operating on the boundary regions.
The strategy for communication between neighboring slices to facilitate these operations is called a \textit{halo-exchange}.
We implement this exchange by padding every slice with buffer regions of \textit{halo size} ($h$) that are shared between the slices on adjacent processes along the mesh. 
The choice of \textit{halo size} depends on two factors\footnote{In case of the pyramid scheme, the halo size is defined on the high resolution grid and we reduce it by the corresponding downsampling factor for the coarse grid.}, apart from the increased memory cost.
Firstly, the halo size should be large enough to ensure an accurate computation of smoothing operations, i.e. larger than half the smoothing kernel size.
Secondly, to keep communications and book keeping to a minimum, in the current implementation particles are not transferred to neighbouring processes if they travel outside of the domain of a given process. This means that we require halo regions to be larger than the expected maximum displacement of a particle over the entire period of the simulation. For large scale cosmological simulations, this displacement is $\sim 20$ Mpc/h in physical size.

\begin{figure}[h]
\centering
\includegraphics[width=.4\textwidth]{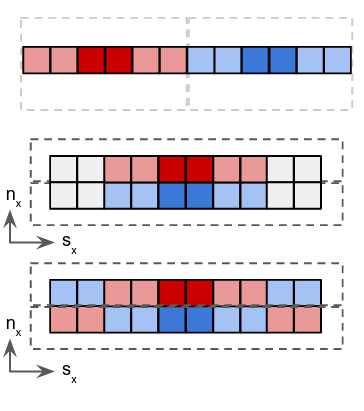}
\caption{An illustration of the halo exchange strategy as outlined in Algorithm \ref{alg:haloex} for a simple 1-D tensor of size $N=12$ and \textit{halo size} $h=2$, specifically $\rm Reshape\_Expand$ operation.
The red and blue regions are to be split on two process, with light regions indicating the overlapping volume that will be exchanged.
In the second panel, the tensor is reshaped with the split along the first, $nx$ dimension and the second un-split $s_x$ that is zero-padded.
In the third panel, the regions are exchanged (here under the assumption of periodic boundary conditions).
}
\label{fig:haloex}
\end{figure}

Our halo-exchange algorithm is outlined in Algorithm \ref{alg:haloex} and illustrated for a 1-dimension grid in Figure \ref{fig:haloex}.
It is primarily motivated by the fact that Mesh-TensorFlow allows to implement independent, local operations simply and efficiently as \textit{slicewise} operations along the dimensions that are not distributed.
Thus we can implement all of the operations - convolutions, CIC interpolations, position and velocity updates for particles and local FFTs - on these un-split dimensions as one would for a typical single-process implementation without worrying about distributed strategies.
To take advantage of this, we begin by reshaping our tensors from 3 dimensions (excluding batch dimension) to 6 dimensions such that the first three indices split the tensor according the computational layout 
and the last three indices are not split, but replicated across all the processes and correspond to the local slice of the grid on each process.
This is shown in Figure \ref{fig:haloex}, where an original 1-D tensor of size $N=12$ in split over $nx=2$ processes.
In the second panel, this is re-shaped to a 2-D tensor with the first dimension $nx=2$ split over processes and $s_x$ is the un-split dimension.
The un-split dimension is then zero-padded with \textit{halo size} resulting in the size along the un-split dimension to be $s_x = N//n_x + 2h$.
This is followed by an exchange with the neighboring slices over the physical volume common to the two adjacent processes.

After the slicewise operation updates the local slices, the same steps are followed in reverse.
The padded halo regions are exchanged again to combine updates in the overlapping volume from adjacent processes.
Then the padded regions are dropped and the tensor is reshaped to the original shape.

\begin{algorithm}
\caption{Halo Exchange}
\label{alg:haloex}
\begin{algorithmic}
\Variables
 \State $N$ - size of the global grid
 \State $h$ - halo size, to be padded
 \State $nx,\ ny,\ nz$ - \#splits along three directions
 \State $sx,\ sy,\ sz$ - size of local slice after split
\State nx, ny, nz - Name of dimension along $nx,\ ny,\ nz$
\State sx, sy, sz - Name of dimension along $sx,\ sy,\ sz$
\EndVariables
\\
\Procedure{Reshape\_Expand}{$G$}
\State{\Call{assert}{G.shape == ($N, N, N$)}}
\State{$G \Leftarrow G.{\rm reshape}(nx, ny, nz, sx, sy, sz)$}
\For{$axis \in$ sx, sy, sz}
    \State{$G \Leftarrow \Call{ZeroPad}{G, size=h, axis=axis}$}
\EndFor
\State{\Return{$G$}}
\EndProcedure
\\
\Procedure{Reshape\_Reduce}{$G$}
\State{\Call{assert}{G.shape == ($nx, ny, nz, sx+2h, xy+2h, sz+2h$)}}
\For{$axis \in$ sx, sy, sz}
    \State{$G \Leftarrow \Call{RemovePad}{G, size=h, axis=axis}$}
\EndFor
\State{$G \Leftarrow G.{\rm reshape}(N, N, N)$}
\State{\Return{$G$}}
\EndProcedure
\\
\Procedure{Exchange}{$G$} \\
\Comment{Add updates in overlapping padded regions from neighboring processes}
\EndProcedure
\\
\Procedure{HaloExchangeOp}{$G, SlicewiseOp$} \\
\Comment{Strategy for operating slicewise op with halo exchange}
\State{\Call{assert}{G.shape == ($N, N, N$)}}
\State{$G \Leftarrow \Call{Reshape\_Expand}{G}$}
\State{$G \Leftarrow \Call{Exchange}{G}$}
\State{$G \Leftarrow \Call{SlicewiseOp}{G}$}
\State{$G \Leftarrow \Call{Exchange}{G}$}
\State{$G \Leftarrow \Call{Reshape\_Reduce}{G}$}
\State{\Return{$G$}}
\EndProcedure
\end{algorithmic}
\end{algorithm}

\subsection{Multi-grid Scheme with Multiresolution Pyramids}
\label{sec:forcepyramids}

To implement the multi-grid scheme with multiresolution pyramids, we need a smoothing and a subsampling operation.
While traditionally smoothing operations are performed in Fourier space in cosmology, here we are restricted to these operations in local, pixel space. However this leads to a tradeoff between the size of the smoothing kernel in pixel space, and how compact the corresponding low pass filter is in Fourier space. Our main consideration is to ensure that for the low-resolution component of the pyramid is sufficiently suppressed at half the Nyquist scale of the original field, so as to be able to downsample this low req field with minimum aliasing. Using larger kernels improves the Fourier drop-off but using larger convolution kernels in pixel space implies an increased computational and memory cost of halo-exchanges to be performed at the boundaries.

In our experiments, we find that a 3-D bspline kernel of order 6 balances these trade-offs well and allows us to achieve sub-percent accuracy, as we show later.
Furthermore in \texttt{FlowPM}, we take advantage of highly efficient TensorFlow ops and implement both, the smoothing and subsampling operations together by constructing a 3D convolution filter with this bspline kernel and convolving the underlying field with stride 2 convolutions.
In our default implementation, we repeat these operations twice and our global mesh is 4x coarser than the higher resolution mesh. 
The full algorithm for this REDUCE operation is outlined in Algorithm \ref{alg:pyramid}.
Based on the discussion in Section \ref{sec:pyramids}, we also outline the reverse operation with EXPAND method.

\begin{algorithm}
\caption{Methods for pyramid scheme}
\label{alg:pyramid}
\begin{algorithmic}
\Procedure{REDUCE}{$H,\ f=2,\ n=6,\ S=2$}\\
\Comment{Downsample field $H$ by convolving with a bspline kernel of order $n$ consecutively $f$ times with stride $S$}
\State{$K \Leftarrow \rm{BSpline(n)}$}
\State{$D \Leftarrow H$}
\For{$i \in 0 \dots f$}
    \State{$D \Leftarrow \Call{Conv3D}{D, K, S}$}
\EndFor
\State{\Return{$D$}}
\EndProcedure
\\
\Procedure{EXPAND}{$D,\ f=2,\ n=6,\ S=2$}\\
\Comment{Upsample field $D$ by convolving with a bspline kernel of order $n$ consecutively $f$ times with stride $S$}
\State{$K \Leftarrow \rm{BSpline(n)}\ x\ 8$}
\State{$H \Leftarrow D$}
\For{$i \in 0 \dots f$}
    \State{$H \Leftarrow \Call{TransposedConv3D}{H, K, S}$}
\EndFor
\State{\Return{$H$}}
\EndProcedure
\end{algorithmic}
\end{algorithm}

Finally we are in the position to outline the multi-grid scheme for force computation.
This is outlined in Algorithm \ref{alg:force}.
Schematically, we begin by reducing the high resolution grid to generate the coarse grid.
Then we estimate the force components in all three directions on both the grids, with the key difference that the coarse grid performs distributed FFT on a global mesh while for the high resolution grid, every process performs local FFTs in parallel on the locally available grids.
To recover the correct total force, we need to combine these long and the short range forces together.
For this, we expand the long-range force grid back to the high resolution.
At the same time, we remove the long-range component from the high-resolution grid with a reduce and expand operation, similar to the band-pass level generated in the Laplacian pyramid in Section \ref{sec:pyramids}.
In the end, we combine these long and short range forces to recover the original force.

Note that in implementing a multi-grid scheme for PM evolution, only the force calculation in the kick step is modified, while the rest of the PM scheme remains the same and operates on the original high-resolution grid.

\begin{algorithm}
\caption{Force computation in Muti-grid pyramid method}
\label{alg:force}
\begin{algorithmic}
\Procedure{ForceGrids}{$G$, mode}\\
\Comment{Estimate the component force grids with 3D FFT}
\State{$\tilde{G}\Leftarrow {\rm FFT3D}(G, mode)$}
\For{$i \in x, y, x$}
    \State{${F}_i \Leftarrow {\rm iFFT3D}(\nabla_i \nabla^{-2}\tilde{G}, mode)$}
\EndFor
\State{\Return{$[{F}_x, {F}_y, {F}_z]$}}
\EndProcedure
\\
\Procedure{Force}{$H,\ f=2,\ n=6,\ S=2$} \\
\Comment{Pyramid scheme to estimate force meshes}
\State{$D \Leftarrow \Call{REDUCE}{H,\ f,\ n,\ S}$}
\State{$F_L \Leftarrow \Call{ForceGrids}{D, mode=distributed}$}
\State{$F_S \Leftarrow \Call{ForceGrids}{H, mode=local}$}
\For{$i \in x, y, x$}
    \State{$F_{i, L} \Leftarrow \Call{EXPAND}{F_{i, L},\ f,\ n,\ S}$}
    \State{$F_{i, l} \Leftarrow \Call{REDUCE}{F_{i, S},\ f,\ n,\ S}$}
    \State{$F_{i, l} \Leftarrow \Call{EXPAND}{F_{i, l},\ f,\ n,\ S}$}
    \State{$F_{i, S} \Leftarrow F_{i, S}-F_{i, l}$}
    \State{$F_i \Leftarrow F_{i, L} + F_{i, S}$}
\EndFor
\State{\Return{$[{F}_x, {F}_y, {F}_z]$}}
\EndProcedure
\end{algorithmic}
\end{algorithm}


\section{Scaling and Numerical Accuracy}
\label{sec:scaling}

In this section, we will discuss the numerical accuracy and scaling of these simulations.
Since the novel aspects of \texttt{FlowPM} are a differentiable Mesh-Tensorflow implementation and multigrid PM scheme, our discussion will center around comparison with the corresponding differentiable python implementation of \texttt{FastPM} which shares the same underlying PM scheme.
We do not discuss the accuracy with respect to high resolution N-Body simulations with correct dynamics, and refer the reader to \cite{Feng2016} for details on such a comparison.
Our tests are performed primarily on GPU nodes on Cori supercomputer at NERSC, with each node hosting 8 NVIDIA V100 ('Volta') GPUs each with with 16 GB HBM2 memory and connected with NVLink interconnect.
We compare our results and scaling with the differentiable python code run on Cori-Haswell nodes. 
This consists of \texttt{FastPM} code for forward modeling,  \texttt{vmad}\footnote{\href{https://github.com/rainwoodman/vmad}{https://github.com/rainwoodman/vmad}} for automated differentiation and \texttt{abopt}\footnote{\href{ }{https://github.com/bccp/abopt}} for optimization.

\subsection{Accuracy of the simulations}

We begin by establishing the accuracy of \texttt{FlowPM} for both, mesh and the pyramid scheme with the corresponding \texttt{FastPM} simulation.
We run both the simulations in a 400 Mpc/h box on 128$^3$ grid for 10 steps from $z=9$ to $z=0$ with force-resolution of 1.
The initial conditions are generated at the 1st order in LPT.
The \texttt{FastPM} simulation is run on a single process while the \texttt{FlowPM} simulations are run on different mesh-splits to validate both the multi-process implementations. 
We use halo size of 16 grids cells for all configurations and find that increasing the halo size increases the accuracy marginally for all configurations.
Moreover, given the more stringent memory constraints of GPU, we run the \texttt{FastPM} simulation with a 64-bit precision while the \texttt{FlowPM} simulations are run with 32-bit precision.

\begin{figure}[h]
\centering
\includegraphics[width=0.45\textwidth]{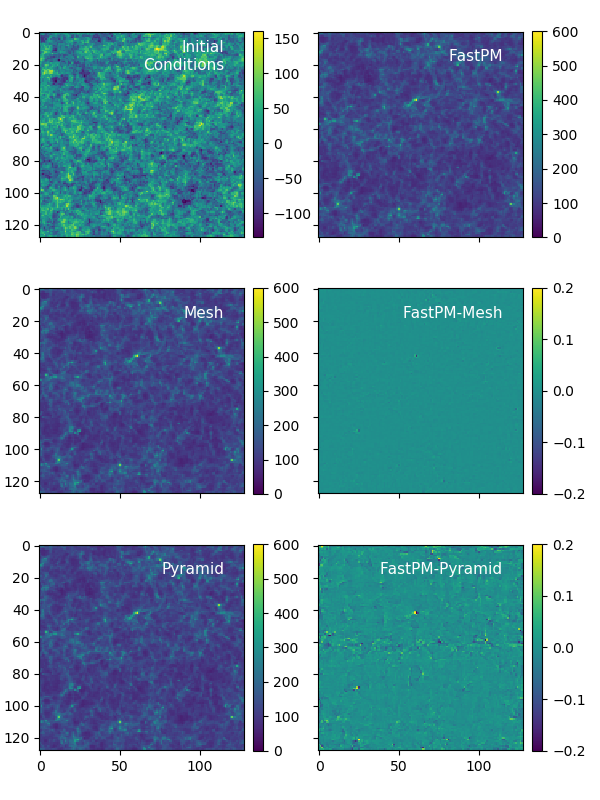}
\caption{Compare the accuracy of \texttt{FlowPM} simulation for Mesh (second left) and Pyramid (third left) scheme with \texttt{FastPM} simulation (top right) for the same initial conditions at the level of fields. The residuals are shown in second and third right image. Configuration is 10 step simulation with 128$^3$ grid and 400 Mpc/h in size. \texttt{FastPM} simulation is run on single process while \texttt{FlowPM} simulations are run on 4 process with nx=2 and ny=2, the splits in x and y direction. Third axis is summed over the box for the projections.}
\label{fig:validim}
\end{figure}

Figure \ref{fig:validim} compares the two simulations at the level of the fields.
For the \texttt{FlowPM} simulations, we show the configuration run on 4 GPUs, with the grid split in 2 along the x and y direction each. 
The pyramid implementation shows higher residuals than the single mesh implementation, but they are sub-percent in either case.

\begin{figure}[h]
\centering
\includegraphics[width=0.45\textwidth]{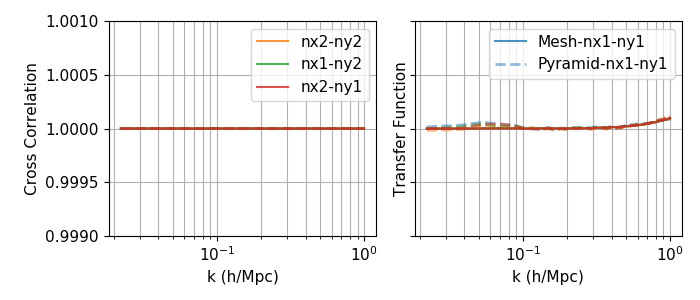}
\caption{Compare the accuracy of \texttt{FlowPM} simulation for Mesh (solid) and Pyramid (dashed) scheme with \texttt{FastPM} simulation at the level of 2 point functions, cross-correlation (left) and transfer-function (right). Configuration is 5 step simulation with 128$^3$ grid and 400 Mpc/h in size. \texttt{FastPM} simulation is run on single process while \texttt{FlowPM} simulations are run on different number of processes with nx and ny as splits in x and y direction respectively, as indicated in the legends.}
\label{fig:valid2pt}
\end{figure}

In Figure \ref{fig:valid2pt} where we compare the two simulations more quantitatively by measuring their clustering. 
To compare their 2-point functions, we measure the transfer function which is the ratio of the power spectra ($P(k)$) of two fields 
\begin{equation}
    T_f (k) = \sqrt{\frac{P_a(k)}{P_b(k)}}
\end{equation}
and their cross correlation which compares the phases and hence is a measure of higher order clustering
\begin{equation}
    r_c (k) = \frac{P_{ab}(k)}{\sqrt{P_a(k) \times P_b(k)}}
\end{equation}
with $P_{ab}$ being the cross power spectra of the two fields. 

Both the transfer function and cross correlation are within $0.01\%$ across all scales, with the latter starting to deviate marginally on small scales. 
Thus the choices made in terms of convolution filters for up-down sampling, halo size and other parameters in multi-grid scheme, though not extensively explored, are adequate to reach the requisite accuracy.
We anticipate this to be a grid resolution dependent statement, and the resolution we have chosen is fairly typical of cosmological simulations that will be run for the analysis of future cosmological surveys and the analytic methods that will most likely use these differentiable simulations. 

\subsection{Scaling on Cori GPU}

We perform the scaling tests on the Cori GPUs for varying grid and mesh sizes, and compare it against the scaling of the python implementation of \texttt{FastPM}.
As mentioned previously, though the accuracy of the simulation is independent of the mesh layout, the performance can be dependent on it.
Thus for the \texttt{FlowPM} implementation, we will look at the timing for different layouts of our mesh.

\begin{figure}[h]
\centering
\includegraphics[width=0.45\textwidth]{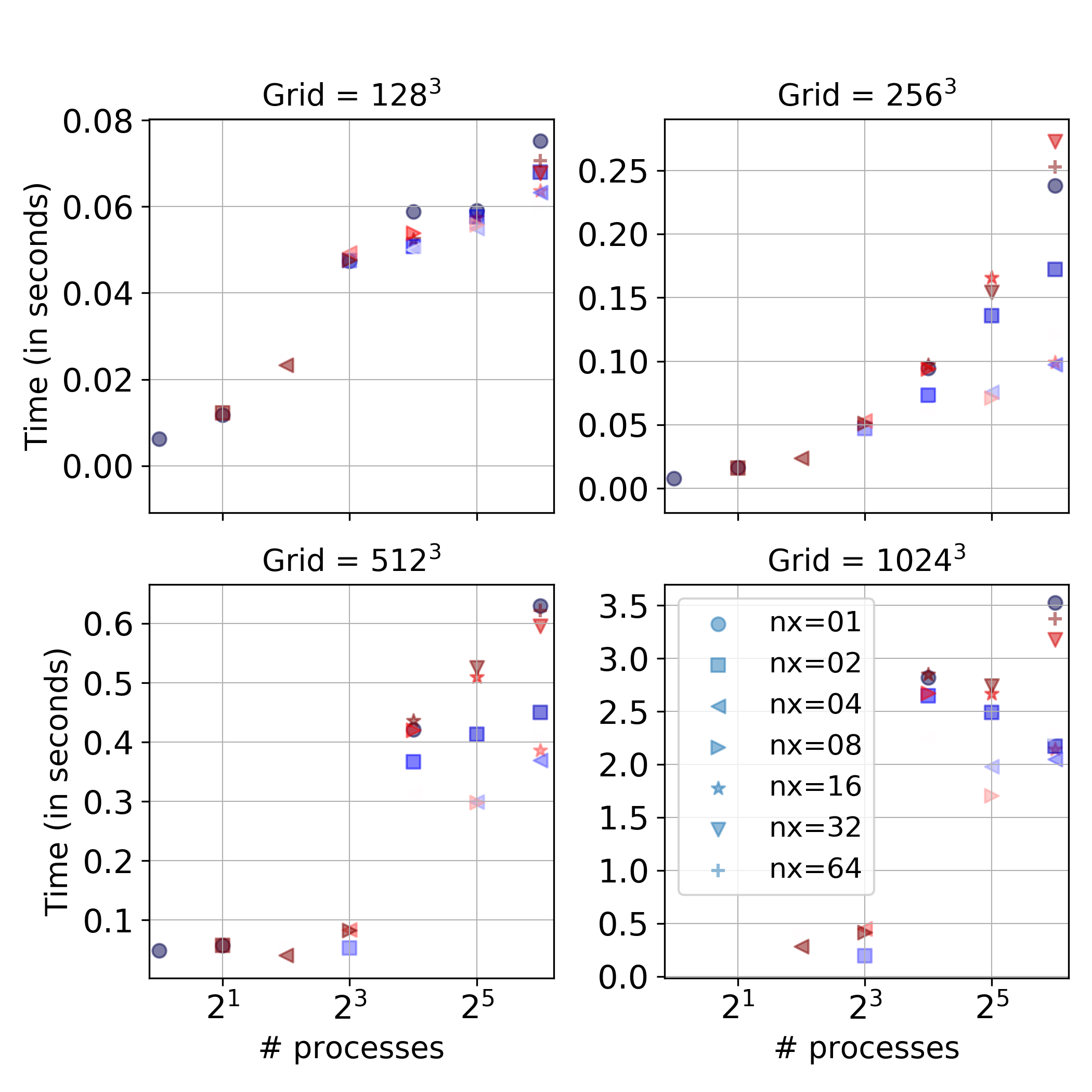}
\caption{Scaling for a single forward+backward 3D FFT implemented in \texttt{FlowPM} in Mesh-Tensorflow for different grid sizes and number of process.
Different symbols indicate different layout for computational mesh, with legends indicating the number of splits along x-direction.
For a given number of processors, we use divergent color-scheme with more splits along x-axis than y-axis shown in red and blue otherwise. We use the same gradient for identical overall configuration.} 
\label{fig:meshfft}
\end{figure}

Since 3D FFTs are typically the most computationally expensive part of a PM simulation, we begin with their time-scaling.
For the Mesh-TF implementation, where we have implemented these operations as low-level Mesh-TensorFlow operators, the time scaling is shown in Figure \ref{fig:meshfft}.
For small grids, the timing for FFT is almost completely dominated by the time spent in all-to-all communications for transpose operations. 
As a result, the scaling is poor with the timing increasing as we increase the number of processes.
However for very large grids, i.e. $N \geq 512$, the compute cost approaches communication cost for small number of process and we see a slight dip upto 8 processes, which is the number of GPUs on a single node.
However after that, further increase in number of processes leads to a significant increase in time due to inter-node communications. 
At the same time, as can be seen from different symbols, unbalanced splits in mesh layout in $x$ and $y$ directions leads to worse performance than a balanced split, with the difference becoming noticeable for large grid sizes.

\begin{figure}[h]
\centering
\includegraphics[width=0.45\textwidth]{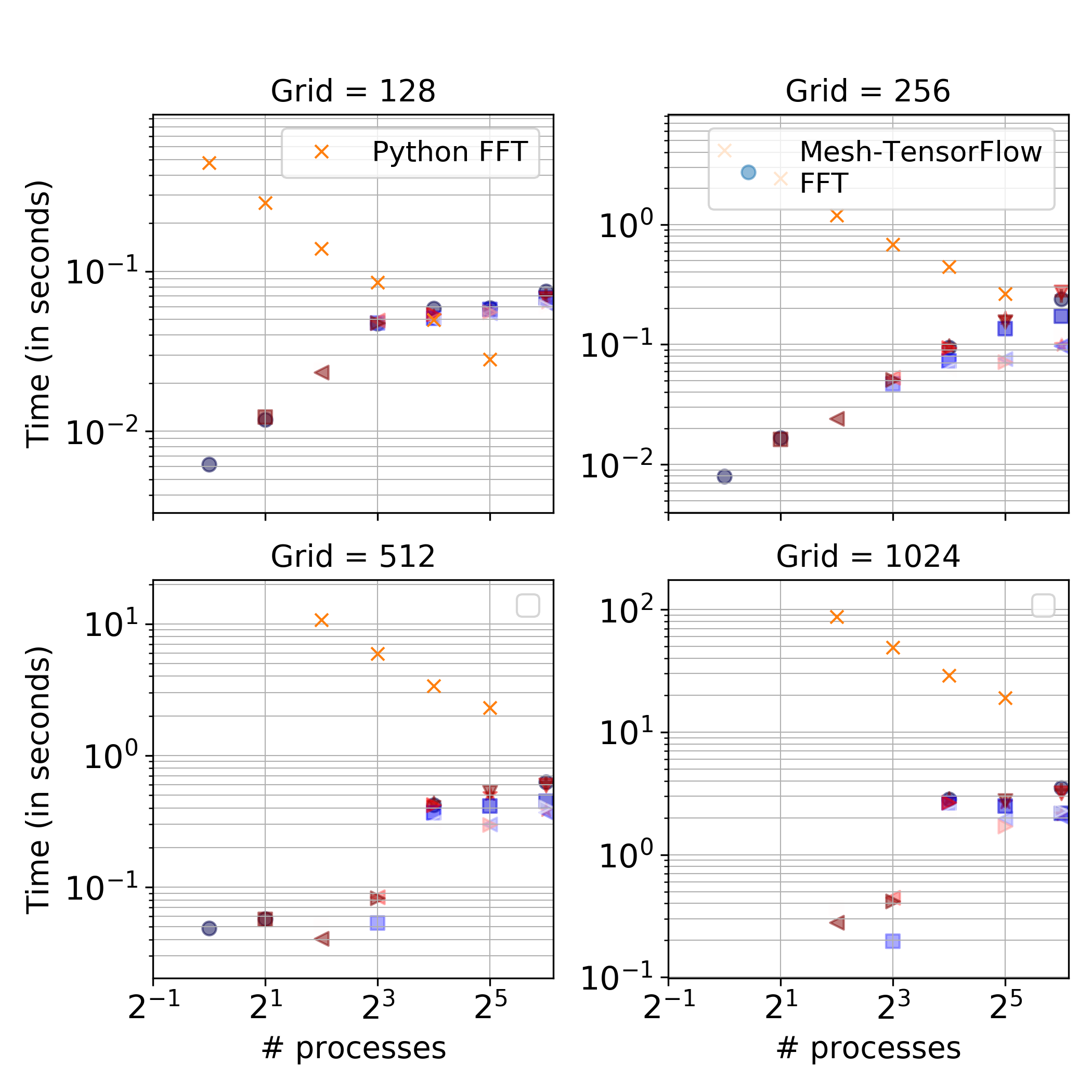}
\caption{Compare the timings of a single forward+backward 3D FFT in \texttt{FlowPM} in Mesh-Tensorflow with PFFT implementation in python-\texttt{FastPM}.
Different layouts for Mesh-TF implementation follow the legend of Figure \ref{fig:meshfft}.}
\label{fig:fftcompare}
\end{figure}

In Figure \ref{fig:fftcompare}, we compare the timing of FFT with that implemented in \texttt{FastPM} which implements the PFFT algorithm for 3D FFTs \citep{pfft}.
The scaling for PFFT is close to linear with the increasing number of processes.
However due to GPU accelerators, the Mesh-TF FFTs are still order of magnitudes faster in most cases, with only getting comparable for small grid sizes and large mesh sizes. 

While the scaling of FFTs is sub-optimal, what is more relevant for us is the scaling of 1 PM step that involves other operations in addition to FFTs.
Thus, next we turn to the scaling of generating initial conditions (ICs) for the cosmological simulation. 
Our ICs are generated as first-order LPT displacements (Zeldovich displacements, \cite{Zeldovich70}).
Schematically, the operations involved in this are outlined in Algorithm \ref{alg:ic}.
It provides a natural test since this step involves all the operations that enter a single PM step i.e. kick, drift, force evaluation and interpolation.

\begin{algorithm}
\caption{Generating IC}
\label{alg:ic}
\begin{algorithmic}
\Procedure{Gen-IC}{$N, pk$}
\State{$g \Leftarrow$ sample 3D normal random field of size $N$}
\State{$\delta_L \Leftarrow$ Scale $g$ by power spectrum $pk$}
\State{$F \Leftarrow$ Estimate force at grid-points from $\delta_L$ i.e. Force}
\State{$d \Leftarrow$ Scale $F$ to obtain displacement}
\State{$X \Leftarrow$ Displace particles with $d$, i.e. Drift}
\State{$V \Leftarrow$ Scale to obtain velocity of particles i.e. Kick}
\State{$\delta \Leftarrow$ Interpolate particles at X to obtain density}
\State{\Return{$\delta$}}
\EndProcedure
\end{algorithmic}
\end{algorithm}

\begin{figure}[h]
\centering
\includegraphics[width=0.45\textwidth]{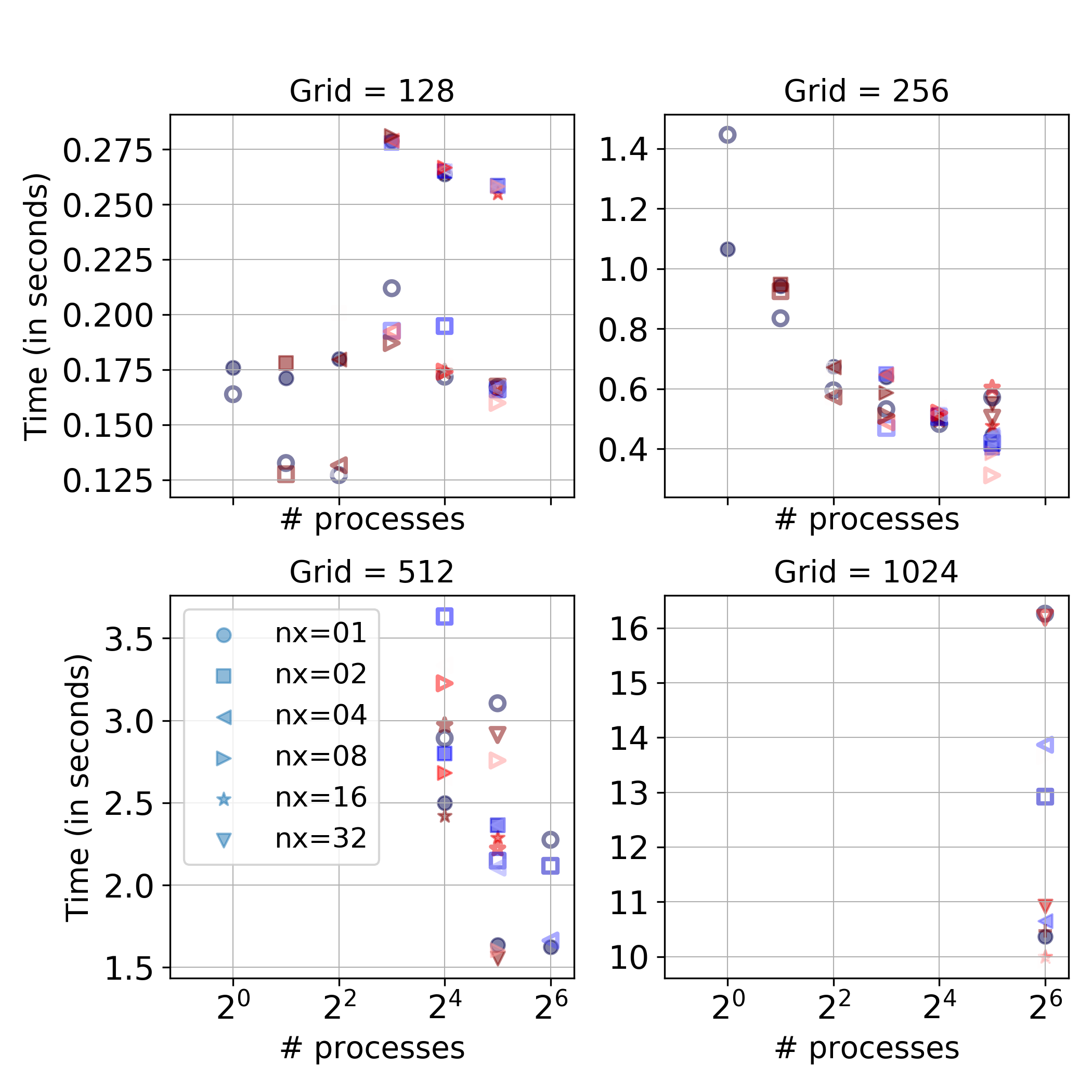}
\caption{Scaling for generating initial conditions with 2nd order LPT in Mesh-Tensorflow for different grid sizes and number of process.
Different symbols indicate different layout for computational mesh, with legends indicating the number of splits along x-direction.
The scaling of pyramid scheme is shown in filled symbols and a single mesh implementation is shown in open symbols. }
\label{fig:meshlpt}
\end{figure}

We establish the scaling for this step in Figure \ref{fig:meshlpt} for both the implementations in \texttt{FlowPM}- the single mesh as well as pyramid scheme. 
Firstly, note that since this step combines computationally intensive operations (like interpolation) with communication intensive operations (like FFT), 
the scaling for a PM step generally has the desirable slope with the timings improving as we increase the number of processes. 
There is, however, still a communication overhead as we move from GPUs on a single node to multi-nodes (see for grid size of 128).
Secondly, as expected, given the increase in number of operations for the pyramid scheme as compared to single-mesh implementation
the single-mesh implementation is more efficient for small grids.
However its scaling is poorer than the pyramid scheme which is $\sim 20\%$ faster for grid size of $N=1024$.

\begin{figure}[h]
\centering
\includegraphics[width=0.45\textwidth]{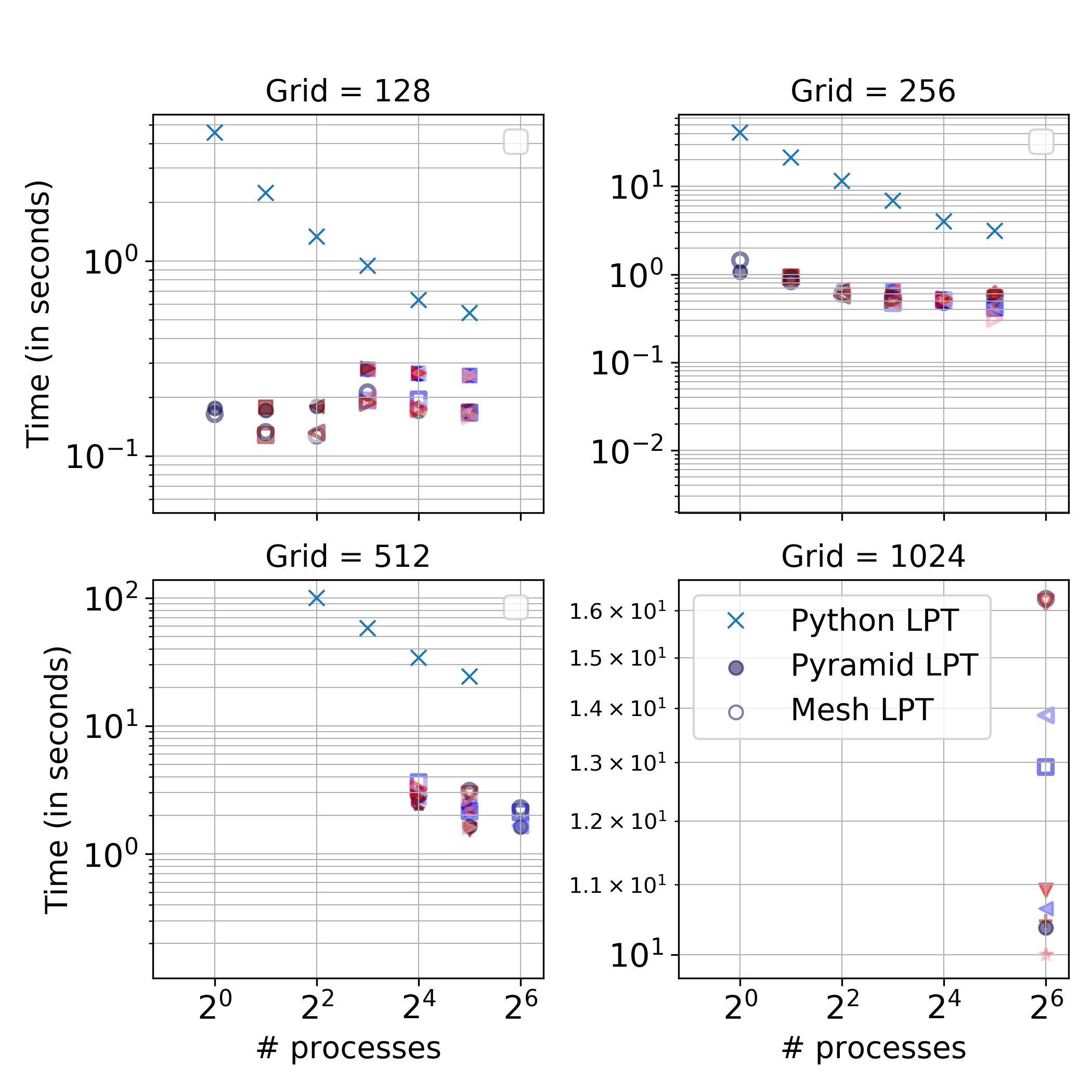}
\caption{Compare the LPT timings for \texttt{FlowPM} in Mesh-Tensorflow and python-\texttt{FastPM}.
Different layouts and schemes for Mesh-TF implementation follow the legend of Figure \ref{fig:meshlpt}.}
\label{fig:lptcompare}
\end{figure}

In Figure \ref{fig:lptcompare}, we compare this implementation with the python implementation.
We find that \texttt{FlowPM} is alteast 10x faster than \texttt{FastPM} for the same number of processes, across all sizes (except the combination of smallest grid and largest mesh size).
Moreover, since the both \texttt{FlowPM} and \texttt{FastPM} benefit with increasing the number of processes, we expect that it will be hard to beat the performance of \texttt{FlowPM} with simply increasing the number of processes in \texttt{FastPM}.

\section{Example application: Reconstruction of initial conditions}
\label{sec:reconstruction}


With the turn of the decade, the next generation of cosmological surveys will probe increasingly smaller scales, over the largest volumes, with different cosmological probes.
As a result, there is a renewed interest in developing forward modeling approaches and simulations based inference techniques to optimally extract and combine information from these surveys. 
Differentiable simulators such as \texttt{FlowPM} will make these approaches tractable in high-dimensional data spaces of cosmology and hence play an important role in the analysis of these future surveys.

Here we demonstrate the efficacy of \texttt{FlowPM} with one such analytic approach that has recently received a lot of attention.
We are interested in reconstructing the initial conditions of the Universe ($\bs$) from the late-time observations ($\bd$) \citep{Wang14, Jasche13, Seljak17, Modi18, Schmittfull19}. 
The late time matter-density field is highly non-Gaussian which makes it hard to model and do inference.
Hence the current analysis methods only extract a fraction of the available information from cosmological surveys.
On the other hand, the initial conditions are known to be Gaussian to a very high degree and hence their power spectrum ($\bS$) is a sufficient statistic.
Thus reconstructing this initial density field provides a way of developing optimal approach for inference \citep{Seljak17}. 
At the same time, the reconstructed initial field can be forward modeled to generate other latent cosmological fields of interest that can be used for supplementary analysis \citep{Modi19, Horowitz19}.

We approach this reconstruction in a forward model Bayesian framework based most directly on \cite{Seljak17, Modi18, Modi19}.
We refer the reader to these for technical details but briefly, 
we forward model ($\mathcal{F}$) the observations of interest from the initial conditions of the Universe ($\bs$) and compare them with the observed data ($\bd$) under a likelihood model - $\mathcal{L}(\bd|\bs)$.
This can be combined with the Gaussian prior on the initial modes to write the corresponding posterior - $\mathcal{P}(\bs|\bd)$. 
This posterior is either optimized to get a maximum-a-posteriori (MAP) estimate of the initial conditions, or alternatively it can be explored with sampling methods as in \cite{Jasche13}. 
However in either approach, given the multi-million dimensionality of the observations, it is necessary to use gradient based approaches for optimization and sampling and hence a differentiable forward model is necessary. 
Particle mesh simulations evolving the initial conditions to the final matter field will form the first part of these forward models for most cosmological observables of interest.
Here we demonstrate in action with two examples how the inbuilt differentiability and interfacing of \texttt{FlowPM} with TensorFlow makes developing such approaches natural.

\subsection{Reconstruction from Dark Matter}

In the first toy problem, consider the reconstruction of the initial matter density field ($\bs$) from the final Eulerian dark matter density field ($\bd = \delta_e$) as an observable. 
While the 3D Eulerian dark matter field is not observed directly in any survey, and weak lensing surveys only allow us to probe the projected matter density field, this problem is illustrative in that the forward model is only the PM simulation i.e. $\mathcal{F}={\rm FlowPM}$ with the modeled density field $\delta_m = \mathcal{F}(\bs)$.
We will include more realistic observation and complex modeling in the next example.

We assume a Gaussian uncorrelated data noise with constant variance $\sigma^2$ in configuration space.
Then, the negative log-posterior of the initial conditions is:
\begin{equation}
\label{eq:posteriordm}
 -\log p(\delta_m | \delta_e) = \frac{(\delta_e - \delta_m )^2}{2 \sigma^2} + \frac{1}{2}\bs^\dagger \bS^{-1}\bs + cst
\end{equation}
where the first term is the negative Gaussian log-likelihood and the second term is the negative Gaussian log-prior on the power spectrum $\bs$ of initial conditions, assuming a fiducial power spectrum $\bS$. 

To reconstruct the initial conditions, we need to minimize the negative log-posterior Eq. \ref{eq:posteriordm}.
The snippet of Mesh-TensorFlow code for such a reconstruction using TF Estimator API \citep{TFesitmator} is outlined in Listing \ref{lst:recondm}.
In the interest of brevity, we have skipped variable names and dimension declarations, data I/O and other \textit{setup code} and included only the logic directly relevant to reconstruction. 
Using Estimators allows us to naturally reuse the entire machinery of TensorFlow including  but not limited to monitoring optimization, choosing from various inbuilt optimization algorithms, restarting optimization from checkpoints and others.  
As with other Estimator APIs, the reconstruction code can be split into three components with minimal functions as -

\begin{figure}[H]
\begin{minted}[frame=single,
               fontsize=\footnotesize]{python}
def recon_model(mesh, data, x0):
    # [...] some code initialisation
    #Define initial conditions as variable
    var=mtf.get_variable(mesh, 
                        'linear', 
                        shape, 
                        tf.constant_initializer(x0))
    #Forward model here with FlowPM
    model=FLOWPM(var, ...)
    #Define metrics for loss
    chisq=mtf.reduce_sum(((model-data)/sigma)^2)
    prior=mtf.reduce_sum(r2c(var)^2/power)
    loss =chisq + prior
    return var, model, loss

def model_fn(x0, data, mode, params):
    # [...] some code initialisation
    var, model, loss=recon_model(mesh, data, x0)
    #Construct optimizer for reconstruction
    if mode == tf.estimator.ModeKeys.TRAIN:
        var_grads=mtf.gradients([loss],[v.outputs
        for v in graph.trainable_variables])
        optimizer=mtf.optimize.AdamOptimizer(0.1)
        update_ops=optimizer.apply_grads(var_grads,
                        graph.trainable_variables)
    #Lower mesh tensorflow variables and ops
    lowering=mtf.Lowering(graph,{mesh:mesh_impl})
    tf_init=lowering.export_to_tf_tensor(var)
    tf_model=lowering.export_to_tf_tensor(model)
    tf_loss=lowering.export_to_tf_tensor(loss)
    #If predict, return current reconstruction
    if mode == tf.estimator.ModeKeys.PREDICT:
        tf.summary.scalar("loss", tf_loss)
        predictions={"ic":tf_init,"data":tf_data}
        return tf.estimator.EstimatorSpec(
          mode=tf.estimator.ModeKeys.PREDICT,
          predictions=predictions)
    #If train, optimize for reconstruction
    if mode == tf.estimator.ModeKeys.TRAIN:
        tf_up_op=[lowering.lowered_operation(op) 
                  for op in update_ops]
        train_op=tf.group(tf_up_op)
        # ...checkpoint hooks...
        return tf.estimator.EstimatorSpec(
                tf.estimator.ModeKeys.TRAIN, 
                loss=tf_loss, train_op=train_op)

def main():
    #  [...] some code initialisation
    #Define estimator for reconstruction
    def input_fn():
        return x0, datafea
    recon_estimator = tf.estimator.Estimator(
            model_fn=model_fn,
            model_dir='./tmp/')
    # Train (Reconstruct)
    recon_estimator.train(input_fn, max_steps=100)
    #and evaluate model.
    eval_results = recon_estimator.predict(input_fn)["ic"]
\end{minted}
\label{lst:recondm}
\caption{(Listing 7.1) Code to reconstruct the initial conditions from the observed matter density field with \texttt{FlowPM} and TensorFlow Estimator API}
\end{figure}

\begin{figure}[H]
\centering
\includegraphics[width=0.45\textwidth]{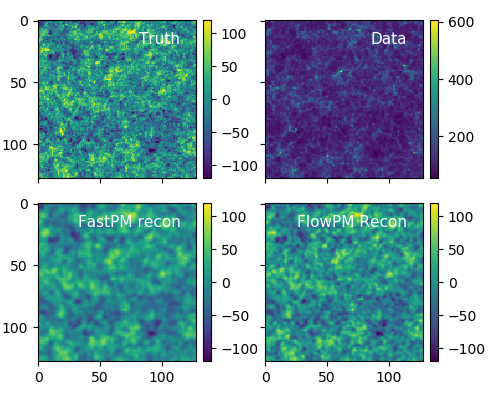}
\caption{Reconstructing initial conditions from the dark matter field observable. Top row shows the true initial conditions (left) and the dark matter data (right). Bottom row shows the reconstructed MAP initial density field with \texttt{FastPM} and FlowPM respectively.
}
\label{fig:recondmim}
\end{figure}

\begin{figure}[H]
\centering
\includegraphics[width=0.45\textwidth]{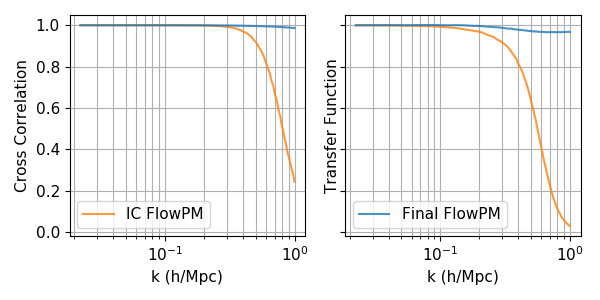}
\caption{Two point statistics for the reconstructed fields from the dark matter field observable. We show the cross correlation coefficient (left) and transfer function (right) for the reconstructed initial field (orange) and the corresponding final dark matter field (blue) with \texttt{FlowPM}. The reconstruction is near perfect on the large scales. 
}
\label{fig:recondm2pt}
\end{figure}

\begin{itemize}
    \item recon\_model : this generates the graph for the forward model using \texttt{FlowPM} from variable linear field (initial conditions) and metrics to optimize for reconstruction 
    \item model\_fn : this creates the train and predict specs for the TF Estimator API, with the train function performing the gradient based updates
    \item main : this calls the estimator to perform and evaluate reconstruction
\end{itemize}
We show the results of this reconstruction in Figure \ref{fig:recondmim} and \ref{fig:recondm2pt}.
Here, the forward model is a 5-step PM simulation in a 400 Mpc/h box on 128$^3$ grid with force resolution of 1.
In addition to gradient based optimization outlined in Listing \ref{lst:recondm}, we implement adiabatic methods as described in \cite{Feng18, Modi18, Modi19} to assist reconstruction of large scales.
We skip those details here in the code for the sake of simplicity, but they are included straightforwardly in this API as a part of recon\_model and training spec.

We compare our results with the previous reconstruction code in python based on \texttt{FastPM}, \texttt{vmad} and \texttt{abopt} (henceforth referred to as \texttt{FastPM} reconstruction).
Both the reconstructions follow the same adiabatic optimization schedule to promote converge on large scales as described in \cite{Modi18, Feng18}.
\texttt{FastPM} reconstruction
uses gradient descent algorithm with single step line-search. Every iteration (including gradient evaluation and line search) takes roughly $\sim\, 4.6$ seconds on 4 nodes i.e. 128 cores on Cori Haswell at NERSC.
We reconstructed for 500 steps, until the large scales converged, in total wall-clock time of $\sim\, 2300$ seconds.
In comparison, for \texttt{FlowPM} reconstruction we take advantage of different algorithms in-built in TensorFlow and use Adam \citep{Kingma2014} optimization with initial learning rate of 0.01, without any early stopping or tolerance.
Every adiabatic optimization step run for 100 iterations (see \cite{Modi18} for details).
With every iteration taking $\sim 1.1$ seconds on a single Cori GPU and 500 iterations for total optimization clocked in at $\sim\, 550$ seconds in wallclock time.
These numbers, except time per iteration, will likely change with a different learning rate and optimization algorithm and we have not explored these in detail for this toy example. 

In Figure \ref{fig:recondmim}, we show the true initial and data field along with the reconstructed initial field by both codes.
In Fig \ref{fig:recondm2pt}, we show the cross correlation and transfer function between the reconstructed and true initial fields (solids) as well as the final data field (dashed) for \texttt{FlowPM}.
\texttt{FastPM} reconstructs the fields comparably, even though we use a different optimization algorithm. 
\texttt{abopt} also provides the flexibility to use L-BFGS algorithm which improves the \texttt{FastPM} reconstruction slightly with a little increase in wall clock of time, up to 4.8 seconds per iteration. 

\subsection{Reconstruction from Halos with a Neural Network model}

In the next example, we consider a realistic case of reconstruction of the initial conditions from dark matter halos under a more complex forward model. 
Dark matter halos and their masses are not observed themselves.
However they are a good proxy for galaxies which are the primary tracers observed in large scale structure surveys, since they capture the challenges posed by galaxies as a discrete, sparse and biased tracer of the underlying matter field. 
Traditionally, halos are modeled as a biased version of the  Eulerian or Lagrangian matter and associated density field  \citep{ST99, Matsubara08b, Schmittfull19}.
To better capture the discrete nature of the observables at the field level, \cite{Modi18} proposed using neural networks to model the halo masses and positions from the Eulerian matter density field. 
Here we replicate their formalism and reconstruct the initial conditions from the halo mass field as observable.

In \cite{Modi18}, the output of PM simulation is supplemented with two pre-trained fully connected networks, NNp and NNm, to predict a mask of halo positions and estimates of halo masses at every grid point. 
These are then combined to predict a halo-mass field ($\rm M_{NN}$) that is compared with the observed halo-mass field ($\rm M_d$) upto a pre-chosen number density.
They propose a heuristic likelihood for the data inspired from the log-normal scatter of halo masses with respect to observed stellar luminosity. Specifically, 

\begin{equation}
    \mathcal{L} = \frac{\mu + \rm{log}(M_d^R + M_0) - \rm{log}(M_{NN}^R + M_0))}{2\sigma^2}    
\end{equation}
where $\rm M_0$ is a constant varied over iterations to assist with optimization, $\rm M^R$ are mass-fields smoothed with a Gaussian of scale $R$ and $\mu$ and $\sigma$ are mass-dependent mean and standard deviations of the log-normal error model estimated from simulations. 

The code snippet in Listing \ref{lst:reconhalo} modifies the code in \ref{lst:recondm} by including these pre-trained neural networks NNp and NNm in the forward model to supplement \texttt{FlowPM}.
It also demonstrates how to include associated variables like $\rm M_0$ in the recon\_model function that can be updated with ease over iterations to modify the model and optimization. 

\begin{figure}[H]
\begin{minted}[frame=single,
               fontsize=\footnotesize]{python}

def recon_model(mesh, data, x0, M0):
    # [...] some code initialisation
    # Load pre-trained variables of NN
    # Define initial conditions as variable
    var = mtf.get_variable(mesh, 'linear', 
            shape, tf.constant_initializer(x0))
    # Forward model here with FlowPM
    final_field = FLOWPM(var, ...)
    # Supplement FlowPM with NN models
    position = NNp(field_field)
    mass = NNm(final_field)
    model = position*mass
    # Define metrics for loss
    chisq = mtf.reduce_sum(
            ((mu + mtf.log(model+M0) -
            mtf.log(data+M0))/sigma)^2)
    prior = mtf.reduce_sum(r2c(var)^2/power)
    loss = chisq + prior
    return var, model, loss

def model_fn(ft, data, mode, params):
    # [...] some code initialisation 
    var, model, loss = recon_model(mesh, data,
            ft['x0'], ft['M0'])
    # same as Figure 12 from here
    
def main():
    # [...] some code initialisation 
    # Estimator with dict input
    def input_fn():
        ft = {'x0':x0, 'M0':M0}
        return ft, data
    # same as Figure 12 from here
    recon_estimator = tf.estimator.Estimator(
            model_fn=model_fn,model_dir=./tmp/)
    # Train (Reconstruct)
    recon_estimator.train(input_fn=input_fn,
            max_steps=100)
    # and evaluate model.
    eval_results = recon_estimator.predict(
            input_fn=input_fn)["ic"]

\end{minted}
\label{lst:reconhalo}
\caption{(Listing 7.2) Update Listing \ref{lst:recondm} to include neural networks in the forward model and associated variables to be changed over iterations of reconstruction}
\end{figure}

\begin{figure}[h]
\centering
\includegraphics[width=0.45\textwidth]{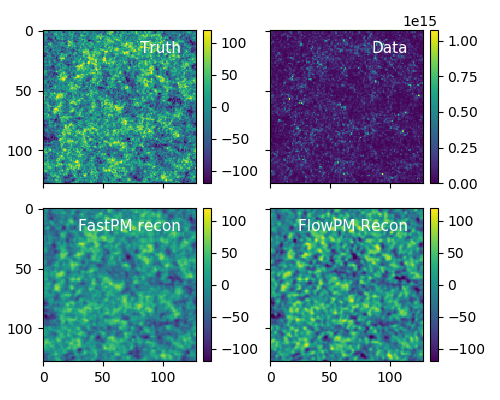}
\caption{Reconstructing initial conditions from the discrete halo mass field observable. Top row shows the true initial conditions (left) and the halo mass field (right). Bottom row shows the reconstructed MAP initial density field with \texttt{FastPM} and FlowPM respectively.}
\label{fig:reconimfnn}
\end{figure}

\begin{figure}[H]
\centering
\includegraphics[width=0.45\textwidth]{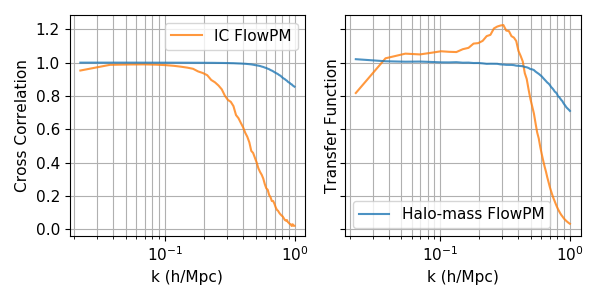}
\caption{Two point statistics for the reconstructed fields from the halo mass field observable. We show the cross correlation coefficient (left) and transfer function (right) for the reconstructed initial field (orange) and the corresponding halo mass field  (blue) with FlowPM. 
}
\label{fig:recon2ptfnn}
\end{figure}

We again compare the results for \texttt{FastPM} and \texttt{FlowPM} reconstruction with the same configuration as before in Figure \ref{fig:reconimfnn} and \ref{fig:recon2ptfnn}.
Here, the data is halo mass field with number density $\bar{n} = 10^{-3}$ (Mpc/h)$^{-3}$ on a 128$^3$ grid for a 400 Mpc/h box.
At this resolution and number density, only $17.5\%$ of grid points are non-zero, indicating the sparsity of our data.
The forward model is a 5 step \texttt{FastPM} simulation followed by two layer fully connected networks with total 5000 and 600 weights respectively (see \cite{Modi18} for details).
The position network takes in non-local features as a flattened array that can also be implemented as a convolution kernel of size 3 in TensorFlow. 
Both the codes reconstruct the initial conditions equally well at the level of cross-correlation.
However the transfer functions for both the reconstruction is quite different.
\cite{Seljak17, Modi18, Horowitz18} discuss simulation based methods to correct the reconstructed transfer function, but we do not discuss them here since it is beyond the scope and not the main point of this work.

In addition to the ease of implementing reconstruction in FlowPM, the most significant gain of \texttt{FlowPM} is in terms of the speed of iteration. 
The time taken for 1 \texttt{FlowPM} iteration is $\sim 1.7$ second while the time taken for 1 iteration in \texttt{FastPM} reconstruction in $\sim 15$ seconds. 
The huge increase in time for \texttt{FastPM} iterations as compared to \texttt{FlowPM} is due to the python implementation of single convolution kernel and its gradients as required by the neural network bias model, which is very efficiently implemented in TensorFlow. 
Due to increased complexity in the forward model, gradient descent with line search does not converge at all.
Thus we use LBFGS optimization for \texttt{FastPM} reconstruction and it takes roughly 450 iterations.
FlowPM implementation, with Adam algorithm as before, does 100 iterations at every step and totals to 1500 iterations since we do not include early stopping.
As a result, \texttt{FlowPM} implementation takes roughly 30 minutes on a single GPU while \texttt{FastPM} reconstruction takes roughly 2 hours on 128 processes. 

\section{Conclusion}
\label{sec:discussion}
In this work, we present \texttt{FlowPM} - a cosmological N-body code implemented in Mesh-TensorFlow for GPU-accelerated, distributed, and differentiable simulations to tackle the unprecedented modeling and analytic challenges posed by the next generation of cosmological surveys. 

\texttt{FlowPM} implements \texttt{FastPM} scheme as the underlying particle-mesh gravity solver, with gravitational force estimated with 3D Fourier transforms.
For distributed computation, we use Mesh-TensorFlow as our model parallelism framework which gives us full flexibility in determining the computational layout and distribution of our simulation.
To overcome the bottleneck of large scale distributed 3D FFTs, we propose and implement a novel multi-grid force computation scheme based on image pyramids.
We demonstrate that without any fine tuning, this method is able to achieve sub-percent accuracy while reducing the communication data-volume by a factor of 64x.
At the same time, given the GPU accelerations, \texttt{FlowPM} is 4-20x faster than the corresponding python \texttt{FastPM} simulation depending on the resolution and compute distribution. 

However the main advantage of \texttt{FlowPM} is the differentiability of the simulation and natural interfacing with deep learning frameworks.
Built entirely in TensorFlow, the simulations are differentiable with respect to every component.
This allows for development of novel analytic methods such as simulations based inference and reconstruction of cosmological fields, that were hitherto intractable due to the multi-million dimensionality of these problems.
We demonstrate with two examples, providing the logical code-snippets, how \texttt{FlowPM} makes the latter straightforward by using TensorFlow Estimator API to do optimization in 128$^3$ dimensional space.
It is also able to naturally interface with machine learning frameworks as a part of the forward model in this reconstruction. 
Lastly, due to its speed, it is able to achieve comparable accuracy to \texttt{FastPM} based reconstruction in 5 times lower wall-clock time, and 640 times lower computing time. 

FlowPM is open-source and publicly available on our Github repo at \href{ }{https://github.com/modichirag/flowpm}.
We provide example code to do forward PM evolution with single-grid and multi-grid force computation.
We also provide code for the reconstruction with both the examples demonstrated in the paper.
We hope that this will encourage the community to use this novel tool and develop scientific methods to tackle the next generation of cosmological surveys.

\section*{Acknowledgements}
We would like to thank Mustafa Mustafa and Wahid Bhimji at NERSC and Thiru Palanisamy at Google for the support in developing this.
We would also like to acknowledge the gracious donation of TPU compute time from Google which was instrumental in developing and testing FlowPM for SIMD Mesh implementation. 
This material is based upon work supported by the National Science Foundation under Grant Numbers 1814370 and NSF 1839217, and by NASA under Grant Number 80NSSC18K1274. This research used resources of the National Energy Research Scientific Computing Center (NERSC), a U.S. Department of Energy Office of Science User Facility operated under Contract No. DE-AC02-05CH11231.


\bibliographystyle{elsarticle-harv} 
\bibliography{main}

\begin{thebibliography}{37}
\expandafter\ifx\csname natexlab\endcsname\relax\def\natexlab#1{#1}\fi
\providecommand{\url}[1]{\texttt{#1}}
\providecommand{\href}[2]{#2}
\providecommand{\path}[1]{#1}
\providecommand{\DOIprefix}{doi:}
\providecommand{\ArXivprefix}{arXiv:}
\providecommand{\URLprefix}{URL: }
\providecommand{\Pubmedprefix}{pmid:}
\providecommand{\doi}[1]{\href{http://dx.doi.org/#1}{\path{#1}}}
\providecommand{\Pubmed}[1]{\href{pmid:#1}{\path{#1}}}
\providecommand{\bibinfo}[2]{#2}
\ifx\xfnm\relax \def\xfnm[#1]{\unskip,\space#1}\fi
\bibitem[{{Alsing} et~al.(2018){Alsing}, {Wandelt} and {Feeney}}]{Alsing18}
\bibinfo{author}{{Alsing}, J.}, \bibinfo{author}{{Wandelt}, B.},
  \bibinfo{author}{{Feeney}, S.}, \bibinfo{year}{2018}.
\newblock \bibinfo{title}{{Massive optimal data compression and density
  estimation for scalable, likelihood-free inference in cosmology}}.
\newblock \bibinfo{journal}{\mnras} \bibinfo{volume}{477},
  \bibinfo{pages}{2874--2885}.
\newblock \DOIprefix\doi{10.1093/mnras/sty819},
  \href{http://arxiv.org/abs/1801.01497}{{\tt arXiv:1801.01497}}.
\bibitem[{Anderson et~al.(1984)Anderson, Bergen, Burt and Ogden}]{Anderson84}
\bibinfo{author}{Anderson, C.H.}, \bibinfo{author}{Bergen, J.R.},
  \bibinfo{author}{Burt, P.J.}, \bibinfo{author}{Ogden, J.M.},
  \bibinfo{year}{1984}.
\newblock \bibinfo{title}{Pyramid methods in image processing}.
\bibitem[{{Burt} and {Adelson}(1983)}]{Burt83}
\bibinfo{author}{{Burt}, P.}, \bibinfo{author}{{Adelson}, E.},
  \bibinfo{year}{1983}.
\newblock \bibinfo{title}{The laplacian pyramid as a compact image code}.
\newblock \bibinfo{journal}{IEEE Transactions on Communications}
  \bibinfo{volume}{31}, \bibinfo{pages}{532--540}.
\bibitem[{{Cheng} et~al.(2017){Cheng}, {Haque}, {Hong}, {Ispir}, {Mewald},
  {Polosukhin}, {Roumpos}, {Sculley}, {Smith}, {Soergel}, {Tang}, {Tucker},
  {Wicke}, {Xia} and {Xie}}]{TFesitmator}
\bibinfo{author}{{Cheng}, H.T.}, \bibinfo{author}{{Haque}, Z.},
  \bibinfo{author}{{Hong}, L.}, \bibinfo{author}{{Ispir}, M.},
  \bibinfo{author}{{Mewald}, C.}, \bibinfo{author}{{Polosukhin}, I.},
  \bibinfo{author}{{Roumpos}, G.}, \bibinfo{author}{{Sculley}, D.},
  \bibinfo{author}{{Smith}, J.}, \bibinfo{author}{{Soergel}, D.},
  \bibinfo{author}{{Tang}, Y.}, \bibinfo{author}{{Tucker}, P.},
  \bibinfo{author}{{Wicke}, M.}, \bibinfo{author}{{Xia}, C.},
  \bibinfo{author}{{Xie}, J.}, \bibinfo{year}{2017}.
\newblock \bibinfo{title}{{TensorFlow Estimators: Managing Simplicity vs.
  Flexibility in High-Level Machine Learning Frameworks}}.
\newblock \bibinfo{journal}{arXiv e-prints} ,
  \bibinfo{pages}{arXiv:1708.02637}\href{http://arxiv.org/abs/1708.02637}{{\tt
  arXiv:1708.02637}}.
\bibitem[{{Cranmer} et~al.(2019){Cranmer}, {Brehmer} and {Louppe}}]{Cranmer19}
\bibinfo{author}{{Cranmer}, K.}, \bibinfo{author}{{Brehmer}, J.},
  \bibinfo{author}{{Louppe}, G.}, \bibinfo{year}{2019}.
\newblock \bibinfo{title}{{The frontier of simulation-based inference}}.
\newblock \bibinfo{journal}{arXiv e-prints} ,
  \bibinfo{pages}{arXiv:1911.01429}\href{http://arxiv.org/abs/1911.01429}{{\tt
  arXiv:1911.01429}}.
\bibitem[{{DESI Collaboration} et~al.(2016){DESI Collaboration}, {Aghamousa},
  {Aguilar}, {Ahlen}, {Alam}, {Allen}, {Allende Prieto}, {Annis}, {Bailey},
  {Balland} and et~al.}]{DESI}
\bibinfo{author}{{DESI Collaboration}}, \bibinfo{author}{{Aghamousa}, A.},
  \bibinfo{author}{{Aguilar}, J.}, \bibinfo{author}{{Ahlen}, S.},
  \bibinfo{author}{{Alam}, S.}, \bibinfo{author}{{Allen}, L.E.},
  \bibinfo{author}{{Allende Prieto}, C.}, \bibinfo{author}{{Annis}, J.},
  \bibinfo{author}{{Bailey}, S.}, \bibinfo{author}{{Balland}, C.},
  \bibinfo{author}{et~al.}, \bibinfo{year}{2016}.
\newblock \bibinfo{title}{{The DESI Experiment Part I: Science,Targeting, and
  Survey Design}}.
\newblock \bibinfo{journal}{ArXiv e-prints}
  \href{http://arxiv.org/abs/1611.00036}{{\tt arXiv:1611.00036}}.
\bibitem[{{Feng} et~al.(2016){Feng}, {Chu}, {Seljak} and {McDonald}}]{Feng2016}
\bibinfo{author}{{Feng}, Y.}, \bibinfo{author}{{Chu}, M.Y.},
  \bibinfo{author}{{Seljak}, U.}, \bibinfo{author}{{McDonald}, P.},
  \bibinfo{year}{2016}.
\newblock \bibinfo{title}{{FASTPM: a new scheme for fast simulations of dark
  matter and haloes}}.
\newblock \bibinfo{journal}{MNRAS} \bibinfo{volume}{463},
  \bibinfo{pages}{2273--2286}.
\newblock \DOIprefix\doi{10.1093/mnras/stw2123},
  \href{http://arxiv.org/abs/1603.00476}{{\tt arXiv:1603.00476}}.
\bibitem[{{Feng} et~al.(2018){Feng}, {Seljak} and {Zaldarriaga}}]{Feng18}
\bibinfo{author}{{Feng}, Y.}, \bibinfo{author}{{Seljak}, U.},
  \bibinfo{author}{{Zaldarriaga}, M.}, \bibinfo{year}{2018}.
\newblock \bibinfo{title}{{Exploring the posterior surface of the large scale
  structure reconstruction}}.
\newblock \bibinfo{journal}{Journal of Cosmology and Astro-Particle Physics}
  \bibinfo{volume}{2018}, \bibinfo{pages}{043}.
\newblock \DOIprefix\doi{10.1088/1475-7516/2018/07/043},
  \href{http://arxiv.org/abs/1804.09687}{{\tt arXiv:1804.09687}}.
\bibitem[{{Greengard} and {Rokhlin}(1987)}]{Greengard87}
\bibinfo{author}{{Greengard}, L.}, \bibinfo{author}{{Rokhlin}, V.},
  \bibinfo{year}{1987}.
\newblock \bibinfo{title}{{A Fast Algorithm for Particle Simulations}}.
\newblock \bibinfo{journal}{Journal of Computational Physics}
  \bibinfo{volume}{73}, \bibinfo{pages}{325--348}.
\newblock \DOIprefix\doi{10.1016/0021-9991(87)90140-9}.
\bibitem[{{Harnois-D{\'e}raps} et~al.(2013){Harnois-D{\'e}raps}, {Pen},
  {Iliev}, {Merz}, {Emberson} and {Desjacques}}]{Harnois13}
\bibinfo{author}{{Harnois-D{\'e}raps}, J.}, \bibinfo{author}{{Pen}, U.L.},
  \bibinfo{author}{{Iliev}, I.T.}, \bibinfo{author}{{Merz}, H.},
  \bibinfo{author}{{Emberson}, J.D.}, \bibinfo{author}{{Desjacques}, V.},
  \bibinfo{year}{2013}.
\newblock \bibinfo{title}{{High-performance P$^{3}$M N-body code:
  CUBEP$^{3}$M}}.
\newblock \bibinfo{journal}{\mnras} \bibinfo{volume}{436},
  \bibinfo{pages}{540--559}.
\newblock \DOIprefix\doi{10.1093/mnras/stt1591},
  \href{http://arxiv.org/abs/1208.5098}{{\tt arXiv:1208.5098}}.
\bibitem[{{He} et~al.(2019){He}, {Li}, {Feng}, {Ho}, {Ravanbakhsh}, {Chen} and
  {P{\'o}czos}}]{He19}
\bibinfo{author}{{He}, S.}, \bibinfo{author}{{Li}, Y.},
  \bibinfo{author}{{Feng}, Y.}, \bibinfo{author}{{Ho}, S.},
  \bibinfo{author}{{Ravanbakhsh}, S.}, \bibinfo{author}{{Chen}, W.},
  \bibinfo{author}{{P{\'o}czos}, B.}, \bibinfo{year}{2019}.
\newblock \bibinfo{title}{{Learning to predict the cosmological structure
  formation}}.
\newblock \bibinfo{journal}{Proceedings of the National Academy of Science}
  \bibinfo{volume}{116}, \bibinfo{pages}{13825--13832}.
\newblock \DOIprefix\doi{10.1073/pnas.1821458116},
  \href{http://arxiv.org/abs/1811.06533}{{\tt arXiv:1811.06533}}.
\bibitem[{Hockney and Eastwood(1988)}]{Hockney88}
\bibinfo{author}{Hockney, R.W.}, \bibinfo{author}{Eastwood, J.W.},
  \bibinfo{year}{1988}.
\newblock \bibinfo{title}{Computer Simulation Using Particles}.
\newblock \bibinfo{publisher}{Taylor \& Francis, Inc.},
  \bibinfo{address}{Bristol, PA, USA}.
\bibitem[{{Horowitz} et~al.(2019a){Horowitz}, {Lee}, {White}, {Krolewski} and
  {Ata}}]{Horowitz19}
\bibinfo{author}{{Horowitz}, B.}, \bibinfo{author}{{Lee}, K.G.},
  \bibinfo{author}{{White}, M.}, \bibinfo{author}{{Krolewski}, A.},
  \bibinfo{author}{{Ata}, M.}, \bibinfo{year}{2019}a.
\newblock \bibinfo{title}{{TARDIS. I. A Constrained Reconstruction Approach to
  Modeling the z {\ensuremath{\sim}} 2.5 Cosmic Web Probed by
  Ly{\ensuremath{\alpha}} Forest Tomography}}.
\newblock \bibinfo{journal}{\apj} \bibinfo{volume}{887}, \bibinfo{pages}{61}.
\newblock \DOIprefix\doi{10.3847/1538-4357/ab4d4c},
  \href{http://arxiv.org/abs/1903.09049}{{\tt arXiv:1903.09049}}.
\bibitem[{{Horowitz} et~al.(2019b){Horowitz}, {Seljak} and
  {Aslanyan}}]{Horowitz18}
\bibinfo{author}{{Horowitz}, B.}, \bibinfo{author}{{Seljak}, U.},
  \bibinfo{author}{{Aslanyan}, G.}, \bibinfo{year}{2019}b.
\newblock \bibinfo{title}{{Efficient optimal reconstruction of linear fields
  and band-powers from cosmological data}}.
\newblock \bibinfo{journal}{\jcap} \bibinfo{volume}{2019},
  \bibinfo{pages}{035}.
\newblock \DOIprefix\doi{10.1088/1475-7516/2019/10/035},
  \href{http://arxiv.org/abs/1810.00503}{{\tt arXiv:1810.00503}}.
\bibitem[{{Izard} et~al.(2016){Izard}, {Crocce} and {Fosalba}}]{Izard16}
\bibinfo{author}{{Izard}, A.}, \bibinfo{author}{{Crocce}, M.},
  \bibinfo{author}{{Fosalba}, P.}, \bibinfo{year}{2016}.
\newblock \bibinfo{title}{{ICE-COLA: towards fast and accurate synthetic galaxy
  catalogues optimizing a quasi-N-body method}}.
\newblock \bibinfo{journal}{\mnras} \bibinfo{volume}{459},
  \bibinfo{pages}{2327--2341}.
\newblock \DOIprefix\doi{10.1093/mnras/stw797},
  \href{http://arxiv.org/abs/1509.04685}{{\tt arXiv:1509.04685}}.
\bibitem[{{Jasche} and {Wandelt}(2013)}]{Jasche13}
\bibinfo{author}{{Jasche}, J.}, \bibinfo{author}{{Wandelt}, B.D.},
  \bibinfo{year}{2013}.
\newblock \bibinfo{title}{{Bayesian physical reconstruction of initial
  conditions from large-scale structure surveys}}.
\newblock \bibinfo{journal}{\mnras} \bibinfo{volume}{432},
  \bibinfo{pages}{894--913}.
\newblock \DOIprefix\doi{10.1093/mnras/stt449},
  \href{http://arxiv.org/abs/1203.3639}{{\tt arXiv:1203.3639}}.
\bibitem[{{Kingma} and {Ba}(2014)}]{Kingma2014}
\bibinfo{author}{{Kingma}, D.P.}, \bibinfo{author}{{Ba}, J.},
  \bibinfo{year}{2014}.
\newblock \bibinfo{title}{{Adam: A Method for Stochastic Optimization}}.
\newblock \bibinfo{journal}{ArXiv e-prints}
  \href{http://arxiv.org/abs/1412.6980}{{\tt arXiv:1412.6980}}.
\bibitem[{{Kitaura} et~al.(2014){Kitaura}, {Yepes} and {Prada}}]{Kitaura14}
\bibinfo{author}{{Kitaura}, F.S.}, \bibinfo{author}{{Yepes}, G.},
  \bibinfo{author}{{Prada}, F.}, \bibinfo{year}{2014}.
\newblock \bibinfo{title}{{Modelling baryon acoustic oscillations with
  perturbation theory and stochastic halo biasing.}}
\newblock \bibinfo{journal}{\mnras} \bibinfo{volume}{439},
  \bibinfo{pages}{L21--L25}.
\newblock \DOIprefix\doi{10.1093/mnrasl/slt172},
  \href{http://arxiv.org/abs/1307.3285}{{\tt arXiv:1307.3285}}.
\bibitem[{{Kodi Ramanah} et~al.(2020){Kodi Ramanah}, {Charnock},
  {Villaescusa-Navarro} and {Wandelt}}]{Ramanah20}
\bibinfo{author}{{Kodi Ramanah}, D.}, \bibinfo{author}{{Charnock}, T.},
  \bibinfo{author}{{Villaescusa-Navarro}, F.}, \bibinfo{author}{{Wandelt},
  B.D.}, \bibinfo{year}{2020}.
\newblock \bibinfo{title}{{Super-resolution emulator of cosmological
  simulations using deep physical models}}.
\newblock \bibinfo{journal}{arXiv e-prints} ,
  \bibinfo{pages}{arXiv:2001.05519}\href{http://arxiv.org/abs/2001.05519}{{\tt
  arXiv:2001.05519}}.
\bibitem[{{LSST Science Collaboration} et~al.(2009){LSST Science
  Collaboration}, {Abell}, {Allison}, {Anderson}, {Andrew}, {Angel}, {Armus},
  {Arnett}, {Asztalos}, {Axelrod} and et~al.}]{LSST}
\bibinfo{author}{{LSST Science Collaboration}}, \bibinfo{author}{{Abell},
  P.A.}, \bibinfo{author}{{Allison}, J.}, \bibinfo{author}{{Anderson}, S.F.},
  \bibinfo{author}{{Andrew}, J.R.}, \bibinfo{author}{{Angel}, J.R.P.},
  \bibinfo{author}{{Armus}, L.}, \bibinfo{author}{{Arnett}, D.},
  \bibinfo{author}{{Asztalos}, S.J.}, \bibinfo{author}{{Axelrod}, T.S.},
  \bibinfo{author}{et~al.}, \bibinfo{year}{2009}.
\newblock \bibinfo{title}{{LSST Science Book, Version 2.0}}.
\newblock \bibinfo{journal}{ArXiv e-prints}
  \href{http://arxiv.org/abs/0912.0201}{{\tt arXiv:0912.0201}}.
\bibitem[{{Matsubara}(2008)}]{Matsubara08b}
\bibinfo{author}{{Matsubara}, T.}, \bibinfo{year}{2008}.
\newblock \bibinfo{title}{{Nonlinear perturbation theory with halo bias and
  redshift-space distortions via the Lagrangian picture}}.
\newblock \bibinfo{journal}{\prd} \bibinfo{volume}{78},
  \bibinfo{pages}{083519}.
\newblock \DOIprefix\doi{10.1103/PhysRevD.78.083519},
  \href{http://arxiv.org/abs/0807.1733}{{\tt arXiv:0807.1733}}.
\bibitem[{{Merz} et~al.(2005){Merz}, {Pen} and {Trac}}]{Merz05}
\bibinfo{author}{{Merz}, H.}, \bibinfo{author}{{Pen}, U.L.},
  \bibinfo{author}{{Trac}, H.}, \bibinfo{year}{2005}.
\newblock \bibinfo{title}{{Towards optimal parallel PM N-body codes: PMFAST}}.
\newblock \bibinfo{journal}{\na} \bibinfo{volume}{10},
  \bibinfo{pages}{393--407}.
\newblock \DOIprefix\doi{10.1016/j.newast.2005.02.001},
  \href{http://arxiv.org/abs/astro-ph/0402443}{{\tt arXiv:astro-ph/0402443}}.
\bibitem[{{Modi} et~al.(2019a){Modi}, {Castorina}, {Feng} and
  {White}}]{HiddenValley19}
\bibinfo{author}{{Modi}, C.}, \bibinfo{author}{{Castorina}, E.},
  \bibinfo{author}{{Feng}, Y.}, \bibinfo{author}{{White}, M.},
  \bibinfo{year}{2019}a.
\newblock \bibinfo{title}{{Intensity mapping with neutral hydrogen and the
  Hidden Valley simulations}}.
\newblock \bibinfo{journal}{arXiv e-prints} ,
  \bibinfo{pages}{arXiv:1904.11923}\href{http://arxiv.org/abs/1904.11923}{{\tt
  arXiv:1904.11923}}.
\bibitem[{{Modi} et~al.(2018){Modi}, {Feng} and {Seljak}}]{Modi18}
\bibinfo{author}{{Modi}, C.}, \bibinfo{author}{{Feng}, Y.},
  \bibinfo{author}{{Seljak}, U.}, \bibinfo{year}{2018}.
\newblock \bibinfo{title}{{Cosmological reconstruction from galaxy light:
  neural network based light-matter connection}}.
\newblock \bibinfo{journal}{Journal of Cosmology and Astro-Particle Physics}
  \bibinfo{volume}{10}, \bibinfo{pages}{028}.
\newblock \DOIprefix\doi{10.1088/1475-7516/2018/10/028},
  \href{http://arxiv.org/abs/1805.02247}{{\tt arXiv:1805.02247}}.
\bibitem[{{Modi} et~al.(2019b){Modi}, {White}, {Slosar} and
  {Castorina}}]{Modi19}
\bibinfo{author}{{Modi}, C.}, \bibinfo{author}{{White}, M.},
  \bibinfo{author}{{Slosar}, A.}, \bibinfo{author}{{Castorina}, E.},
  \bibinfo{year}{2019}b.
\newblock \bibinfo{title}{{Reconstructing large-scale structure with neutral
  hydrogen surveys}}.
\newblock \bibinfo{journal}{arXiv e-prints} ,
  \bibinfo{pages}{arXiv:1907.02330}\href{http://arxiv.org/abs/1907.02330}{{\tt
  arXiv:1907.02330}}.
\bibitem[{Pippig(2013)}]{pfft}
\bibinfo{author}{Pippig, M.}, \bibinfo{year}{2013}.
\newblock \bibinfo{title}{Pfft - an extension of fftw to massively parallel
  architectures}.
\newblock \bibinfo{journal}{SIAM Journal on Scientific Computing}
  \bibinfo{volume}{35}, \bibinfo{pages}{C213 -- C236}.
\newblock \DOIprefix\doi{10.1137/120885887}.
\bibitem[{{Quinn} et~al.(1997){Quinn}, {Katz}, {Stadel} and {Lake}}]{Quinn97}
\bibinfo{author}{{Quinn}, T.}, \bibinfo{author}{{Katz}, N.},
  \bibinfo{author}{{Stadel}, J.}, \bibinfo{author}{{Lake}, G.},
  \bibinfo{year}{1997}.
\newblock \bibinfo{title}{{Time stepping N-body simulations}}.
\newblock \bibinfo{journal}{arXiv e-prints} ,
  \bibinfo{pages}{astro--ph/9710043}\href{http://arxiv.org/abs/astro-ph/9710043}{{\tt
  arXiv:astro-ph/9710043}}.
\bibitem[{{Schmittfull} et~al.(2018){Schmittfull}, {Simonovi{\'c}}, {Assassi}
  and {Zaldarriaga}}]{Schmittfull19}
\bibinfo{author}{{Schmittfull}, M.}, \bibinfo{author}{{Simonovi{\'c}}, M.},
  \bibinfo{author}{{Assassi}, V.}, \bibinfo{author}{{Zaldarriaga}, M.},
  \bibinfo{year}{2018}.
\newblock \bibinfo{title}{{Modeling Biased Tracers at the Field Level}}.
\newblock \bibinfo{journal}{arXiv e-prints}
  \href{http://arxiv.org/abs/1811.10640}{{\tt arXiv:1811.10640}}.
\bibitem[{{Seljak} et~al.(2017){Seljak}, {Aslanyan}, {Feng} and
  {Modi}}]{Seljak17}
\bibinfo{author}{{Seljak}, U.}, \bibinfo{author}{{Aslanyan}, G.},
  \bibinfo{author}{{Feng}, Y.}, \bibinfo{author}{{Modi}, C.},
  \bibinfo{year}{2017}.
\newblock \bibinfo{title}{{Towards optimal extraction of cosmological
  information from nonlinear data}}.
\newblock \bibinfo{journal}{Journal of Cosmology and Astro-Particle Physics}
  \bibinfo{volume}{2017}, \bibinfo{pages}{009}.
\newblock \DOIprefix\doi{10.1088/1475-7516/2017/12/009},
  \href{http://arxiv.org/abs/1706.06645}{{\tt arXiv:1706.06645}}.
\bibitem[{Shazeer et~al.(2018)Shazeer, Cheng, Parmar, Tran, Vaswani,
  Koanantakool, Hawkins, Lee, Hong, Young, Sepassi and Hechtman}]{Shazeer2018}
\bibinfo{author}{Shazeer, N.}, \bibinfo{author}{Cheng, Y.},
  \bibinfo{author}{Parmar, N.}, \bibinfo{author}{Tran, D.},
  \bibinfo{author}{Vaswani, A.}, \bibinfo{author}{Koanantakool, P.},
  \bibinfo{author}{Hawkins, P.}, \bibinfo{author}{Lee, H.},
  \bibinfo{author}{Hong, M.}, \bibinfo{author}{Young, C.},
  \bibinfo{author}{Sepassi, R.}, \bibinfo{author}{Hechtman, B.A.},
  \bibinfo{year}{2018}.
\newblock \bibinfo{title}{Mesh-tensorflow: Deep learning for supercomputers}.
\newblock \bibinfo{journal}{CoRR} \bibinfo{volume}{abs/1811.02084}.
\newblock \URLprefix \url{http://arxiv.org/abs/1811.02084},
  \href{http://arxiv.org/abs/1811.02084}{{\tt arXiv:1811.02084}}.
\bibitem[{{Sheth} and {Tormen}(1999)}]{ST99}
\bibinfo{author}{{Sheth}, R.K.}, \bibinfo{author}{{Tormen}, G.},
  \bibinfo{year}{1999}.
\newblock \bibinfo{title}{{Large-scale bias and the peak background split}}.
\newblock \bibinfo{journal}{\mnras} \bibinfo{volume}{308},
  \bibinfo{pages}{119--126}.
\newblock \DOIprefix\doi{10.1046/j.1365-8711.1999.02692.x},
  \href{http://arxiv.org/abs/astro-ph/9901122}{{\tt arXiv:astro-ph/9901122}}.
\bibitem[{{Stein} et~al.(2019){Stein}, {Alvarez} and {Bond}}]{Stein19}
\bibinfo{author}{{Stein}, G.}, \bibinfo{author}{{Alvarez}, M.A.},
  \bibinfo{author}{{Bond}, J.R.}, \bibinfo{year}{2019}.
\newblock \bibinfo{title}{{The mass-Peak Patch algorithm for fast generation of
  deep all-sky dark matter halo catalogues and its N-body validation}}.
\newblock \bibinfo{journal}{\mnras} \bibinfo{volume}{483},
  \bibinfo{pages}{2236--2250}.
\newblock \DOIprefix\doi{10.1093/mnras/sty3226},
  \href{http://arxiv.org/abs/1810.07727}{{\tt arXiv:1810.07727}}.
\bibitem[{{Suisalu} and {Saar}(1995)}]{Suisalu95}
\bibinfo{author}{{Suisalu}, I.}, \bibinfo{author}{{Saar}, E.},
  \bibinfo{year}{1995}.
\newblock \bibinfo{title}{{An adaptive multigrid solver for high-resolution
  cosmological simulations}}.
\newblock \bibinfo{journal}{\mnras} \bibinfo{volume}{274},
  \bibinfo{pages}{287--299}.
\newblock \DOIprefix\doi{10.1093/mnras/274.1.287},
  \href{http://arxiv.org/abs/astro-ph/9412043}{{\tt arXiv:astro-ph/9412043}}.
\bibitem[{{Tassev} et~al.(2013){Tassev}, {Zaldarriaga} and
  {Eisenstein}}]{Tassev13}
\bibinfo{author}{{Tassev}, S.}, \bibinfo{author}{{Zaldarriaga}, M.},
  \bibinfo{author}{{Eisenstein}, D.J.}, \bibinfo{year}{2013}.
\newblock \bibinfo{title}{{Solving large scale structure in ten easy steps with
  COLA}}.
\newblock \bibinfo{journal}{\jcap} \bibinfo{volume}{2013},
  \bibinfo{pages}{036}.
\newblock \DOIprefix\doi{10.1088/1475-7516/2013/06/036},
  \href{http://arxiv.org/abs/1301.0322}{{\tt arXiv:1301.0322}}.
\bibitem[{{Wang} et~al.(2014){Wang}, {Mo}, {Yang}, {Jing} and {Lin}}]{Wang14}
\bibinfo{author}{{Wang}, H.}, \bibinfo{author}{{Mo}, H.J.},
  \bibinfo{author}{{Yang}, X.}, \bibinfo{author}{{Jing}, Y.P.},
  \bibinfo{author}{{Lin}, W.P.}, \bibinfo{year}{2014}.
\newblock \bibinfo{title}{{ELUCID{\textemdash}Exploring the Local Universe with
  the Reconstructed Initial Density Field. I. Hamiltonian Markov Chain Monte
  Carlo Method with Particle Mesh Dynamics}}.
\newblock \bibinfo{journal}{\apj} \bibinfo{volume}{794}, \bibinfo{pages}{94}.
\newblock \DOIprefix\doi{10.1088/0004-637X/794/1/94},
  \href{http://arxiv.org/abs/1407.3451}{{\tt arXiv:1407.3451}}.
\bibitem[{White et~al.(2014)White, Tinker and McBride}]{White14}
\bibinfo{author}{White, M.}, \bibinfo{author}{Tinker, J.L.},
  \bibinfo{author}{McBride, C.K.}, \bibinfo{year}{2014}.
\newblock \bibinfo{title}{Mock galaxy catalogues using the quick particle mesh
  method}.
\bibitem[{{Zel'Dovich}(1970)}]{Zeldovich70}
\bibinfo{author}{{Zel'Dovich}, Y.B.}, \bibinfo{year}{1970}.
\newblock \bibinfo{title}{{Reprint of 1970A\&amp;A.....5...84Z. Gravitational
  instability: an approximate theory for large density perturbations.}}
\newblock \bibinfo{journal}{\aap} \bibinfo{volume}{500},
  \bibinfo{pages}{13--18}.

\end{thebibliography}

\end{document}